\documentclass[a4paper,11pt]{article}
\pdfoutput=1
\usepackage{jcapmod} 
\usepackage[scaled]{helvet}
\usepackage[toc,page]{appendix}
\usepackage{mathtools}
\usepackage{booktabs}
\usepackage[english]{babel}
\usepackage{amsmath,amssymb,amsbsy,amstext, amsthm, simplewick, amsfonts}
\usepackage{hyperref}
\usepackage{graphicx}
\usepackage[small]{caption}
\usepackage{siunitx}
\usepackage{upgreek}
\usepackage{framed}
\usepackage{wrapfig}
\usepackage{multirow}
\usepackage{bbm}
\usepackage[svgnames,dvipsnames,x11names]{xcolor}
\usepackage[utf8x]{inputenc}
\usepackage{selinput}
\usepackage{bm}
\usepackage{float}
\usepackage{geometry}
\usepackage{yfonts}
\usepackage{subcaption}
\usepackage{sidecap}
\usepackage{longtable}
\usepackage{anyfontsize}
\usepackage{dsfont}
\usepackage{tikz}
\usepackage{relsize}
\usepackage{tcolorbox}
\usepackage{shuffle}
\usetikzlibrary{matrix}
\usepackage{bigstrut}
\usepackage{fnpct}
\usetikzlibrary{arrows.meta}
\usetikzlibrary{decorations.pathreplacing,calligraphy}
\usepackage{tensor}
\usepackage{mleftright}

\usetikzlibrary{decorations.markings, decorations.pathmorphing, decorations.pathreplacing, decorations.shapes}

\usepackage{xcolor}
\usepackage{xparse}

\usepackage{fontawesome5}

\makeatletter
\newcommand{\github}[1]{%
   \href{#1}{\faGithub}%
}
\makeatother

\NewDocumentCommand{\colornucleus}{omme{_^}}{%
  \begingroup\colorlet{currcolor}{.}%
  \IfValueTF{#1}
   {\textcolor[#1]{#2}}
   {\textcolor{#2}}
    {%
     #3
     \IfValueT{#4}{_{\textcolor{currcolor}{#4}}}
     \IfValueT{#5}{^{\textcolor{currcolor}{#5}}}
    }%
  \endgroup
}

\usepackage{array}
\newcolumntype{L}[1]{>{\raggedright\let\newline\\\arraybackslash\hspace{0pt}}m{#1}}
\newcolumntype{C}[1]{>{\centering\let\newline\\\arraybackslash\hspace{0pt}}m{#1}}
\newcolumntype{R}[1]{>{\raggedleft\let\newline\\\arraybackslash\hspace{0pt}}m{#1}}

\usepackage[framemethod=default]{mdframed}
\newmdenv[skipabove=7pt,
skipbelow=7pt,
rightline=true,
leftline=true,
topline=true,
bottomline=true,
backgroundcolor=gray!10,
linecolor=black,
innerleftmargin=5pt,
innerrightmargin=5pt,
innertopmargin=5pt,
innerbottommargin=5pt,
leftmargin=0cm,
rightmargin=0cm,
linewidth=1pt]{eBox}

\definecolor{Red}{RGB}{214, 39, 40}
\definecolor{Blue}{RGB} {31, 119, 180}
\definecolor{Orange}{RGB}{255, 153, 51}
\definecolor{Purple}{RGB}{178, 102, 255}
\definecolor{Green}{RGB}{44, 160, 44}
\definecolor{regal}{RGB}{90,0,120}
\definecolor{darkblue}{rgb}{0.15,0.35,0.55}
\definecolor{reddish}{rgb}{0.65, 0.2, 0.2}
\definecolor{darkgreen}{RGB}{50,150,0}
\definecolor{greyish}{rgb}{.90,.90,.90}
\definecolor{greyish2}{rgb}{.96,.96,.96}
\definecolor{greyish3}{rgb}{.37,.37,.37}
\definecolor{darkblue2}{rgb}{0.3,0.4,0.9}
\definecolor{Blue3}{RGB}{31, 119, 180}

\definecolor{blue3}{RGB}{31, 119, 180}
\definecolor{red3}{RGB}{	214, 39, 40}
\definecolor{orange3}{RGB}{255, 127, 14}
\definecolor{green3}{RGB}{44, 160, 44}
\definecolor{repBlue}{RGB}{31, 119, 180}
\definecolor{repRed}{RGB}{	214, 39, 40}
\definecolor{repGreen}{RGB}{44, 160, 44}

\definecolor{vio}{RGB}{19, 130, 164}
\definecolor{vioo}{RGB}{89, 2, 155}
\newcommand{\Comment}[1]{{}}
\usepackage{empheq}

\usepackage[linktocpage=true]{hyperref}
\hypersetup{
colorlinks=true,
citecolor=darkblue,
linkcolor=reddish,
urlcolor=darkblue,
pdfauthor={},
pdftitle={},
pdfsubject={}
}

\newcommand{\dint}{\text{d}}

\usepackage{colortbl}
\definecolor{lightgreen}{cmyk}{0.2, 0, 0.2, 0.2}
\definecolor{lightgray2}{cmyk}{0.1,0.1,0,0.1}
\definecolor{Red2}{RGB}{214, 39, 40}
\definecolor{Blue2}{RGB} {31, 119, 180}
\definecolor{Orange2}{RGB}{255, 127, 14}
\definecolor{Green2}{RGB}{44, 160, 44}

\makeatletter
\newlength{\apb@width}
\newcommand{\autoparbox}[2][c]{\settowidth{\apb@width}{#2}\parbox[#1]{\apb@width}{#2}}

\makeatother


\def\beq{\begin{equation}}
\def\eeq{\end{equation}}
\def\be{\begin{equation}}
\def\ee{\end{equation}}

\DeclarePairedDelimiter\abs{\lvert}{\rvert}

\allowdisplaybreaks[1]
\newcommand\sqmatrix[2][c]{%
  \fixTABwidth{T}%
  \setbox0=\hbox{$\tabbedCenterstack{#2}$}%
  \setstackgap{L}{\dimexpr\maxTAB@width+\tabbed@gap}%
  \tabbedCenterstack[#1]{#2}%
}

\usetikzlibrary{shapes.misc}

\tikzset{cross/.style={cross out, draw=black, minimum size=2*(#1-\pgflinewidth), inner sep=0pt, outer sep=0pt},
cross/.default={1pt}}

\usetikzlibrary{calc}

\begin{document}


\newgeometry{top=2cm, bottom=2cm, left=2cm, right=2cm}

\begin{titlepage}
\setcounter{page}{1} \baselineskip=15.5pt 
\thispagestyle{empty}

\begin{center}
{\fontsize{18}{18} \bf Strongly Coupled Sectors in Inflation:}\vskip 4pt
{\fontsize{14}{14} \bf Gapped Theories of Unparticles}
\end{center}

\vskip 30pt
\begin{center}
\noindent
{\fontsize{12}{18}\selectfont Yikun Jiang,$^1$ Guilherme L.~Pimentel$^{2,3}$ and Chen Yang$^{2,3}$}
\end{center}

\begin{center}
\vskip 4pt
\textit{$^1$ Northeastern University, 360 Huntington Avenue, Boston, MA 02115, USA}
\vskip 4pt
\textit{$^2$ Scuola Normale Superiore, Piazza dei Cavalieri 7, 56126, Pisa, Italy}
\vskip 4pt
\textit{$^3$ INFN-Sezione di Pisa, Largo Pontecorvo 3, Pisa, 56127, Italy}
\end{center}

\vspace{0.4cm}
\begin{center}{\bf Abstract}\end{center}
We consider a novel scenario for a strongly coupled spectator sector during inflation, that of a higher dimensional conformal field theory with large anomalous dimensions---``unparticles"---and compactify the extra dimensions. 
More specifically, we take generalized free fields in five dimensions, where the extra dimension is compactified to a circle. 
Due to the usual Kaluza-Klein mechanism, the resulting excitations carry properties of both particles and unparticles, so we dub this scenario ``gapped unparticles". 
We derive a two-point function of the gapped unparticles by performing dimensional reduction. 
We then compute, in the collapsed limit, the four-point correlation function of conformally coupled scalars exchanging a gapped unparticle, which are used as seed functions to obtain the correlation function of primordial density perturbations. 
The phenomenology of the resulting correlators presents some novel features, such as oscillations with an envelope controlled by the anomalous dimension, rather than the usual value of $3/2$.
Depending on the value of the five-dimensional scaling dimension and effective mass of the gapped unparticles, we find a clear strategy to distinguish gapped unparticles from heavy massive scalars. 
If we assume the interactions are localized on a brane, gapped unparticles with different effective masses will share a universal coupling, and their exchanges produce an interesting interference pattern. 
\end{titlepage}
\restoregeometry

\newpage
\setcounter{tocdepth}{2}
\setcounter{page}{2}

\linespread{0.75}
\tableofcontents
\linespread{1.}

\newpage
\section{Introduction}
\label{sec:intro}

Inflationary model building is often phrased in terms of a slow-roll potential, together with additional spectator fields. A more model-agnostic perspective is to classify the possible ultraviolet (UV) completions of the effective field theory (EFT) of inflation \cite{Cheung:2007st,Baumann:2009ds,Lee:2016vti}. 
In that context, standard slow-roll scenarios are a UV completion akin to the linear sigma model (LSM), where the inflationary background evolution spontaneously breaks the time-translation symmetry, and the adiabatic fluctuations of the background history give rise to the Goldstone boson $\pi$, which sources curvature fluctuations. 
Most models of inflation fall within this scenario, perhaps with additional weakly coupled spectator fields \cite{Maldacena:2002vr,Chen:2010xka,Arkani-Hamed:2015bza,Arkani-Hamed:2018kmz}. 
Another natural UV completion is to regard the spectator fields as strongly coupled composites, made up of more elementary particles. 
This scenario is similar to the models of Composite Higgs \cite{Weinberg:1975gm,Susskind:1978ms,Kaplan:1983fs,Kaplan:1983sm,Dugan:1984hq,Barbieri:2007bh,Bellazzini:2014yua,Panico:2015jxa,Goertz:2018dyw} and Quantum Chromodynamics (QCD).   
This is much less explored in the literature.\footnote{It would also be interesting to have the entire inflationary dynamics being driven by a microscopic theory in which all excitations are composite. We consider the more modest scenario of a spectator sector with strongly coupled dynamics.}
If we parameterize the space of strongly coupled models using the mass gap $\mathrm{M}_{\mathrm{Gap}}$ of the theory, we can divide the possibilities into three major scenarios~\cite{Pimentel:2025rds}: 
\begin{itemize}
    \item $\mathrm{M}_{\mathrm{Gap}}/H\ll1$: when the mass gap is very small compared to the Hubble constant during inflation, we can approximately treat the theory as a conformal field theory (CFT). 
    Some attempts have been made in this scenario, including studies of gapless unparticles \cite{Green:2013rd,Baumgart:2021ptt,Pimentel:2025rds,Yang:2025apy}, continuum isocurvaton \cite{Aoki:2023tjm}, and light compact scalars \cite{Chakraborty:2023eoq,Chakraborty:2025myb,Chakraborty:2025mhh}. 
    These works have focused on the gapless CFTs. 
    \item $\mathrm{M}_{\mathrm{Gap}}/H\sim1$: we can treat inflation as a particle detector for masses below, at and slightly above the Hubble scale. 
    In this scenario, it is possible to produce several bound states of the strongly coupled sector, with masses of order Hubble to propagate for a few Hubble volumes before their power gets diluted. 
    If the modes share a universal coupling, they will interfere, potentially giving a novel imprint on cosmological correlators. 
    \item $\mathrm{M}_{\mathrm{Gap}}/H\gg1$: since the gap is too large compared to the working energy scale ($\sim H$), the imprint of the strongly coupled sector will degenerate with the tree-level exchange of a massive scalar, making it difficult to distinguish this scenario from weakly coupled UV completions \footnote{A possible way out is to have a large density of states near $\mathrm{M}_{\mathrm{Gap}}$.}. 
\end{itemize}
The scope of this work is to inspect the cosmological imprints of a strongly coupled sector in the gapped phase during inflation, including the second and third scenarios. 
A few works discussed the inflationary consequences of gapped strongly coupled sectors, e.g., \cite{Kumar:2018jxz,Kumar:2025anx,Mishra:2025ofh}, while we will focus on gapped unparticles, containing information on both mass gaps and anomalous dimensions. 
Most importantly, we will show that strongly coupled sectors in the gapped phase will leave distinguishable signatures in the collapsed limit of trispectra, which is the kinematic regime in which the imprints of new physics are often most evident (see for example  \cite{Chen:2009we,Chen:2009zp,Baumann:2011nk,Assassi:2012zq,Noumi:2012vr,Cui:2021iie,Reece:2022soh,Qin:2022lva,Pimentel:2022fsc,Wang:2022eop,Wang:2025qww,deRham:2025mjh,Cai:2025kcd,Wang:2025qfh}, and many more). 
We will focus on the collapsed limit of trispectra, because the collapsed limit encodes the main features due to the exchange of new states, thus encapsulating the physics of the spectator sector. 

Extra dimensions, as one of the most studied directions in Beyond the Standard Model (BSM) physics, provides a framework in which to gap unparticles with large anomalous dimensions in a calculable way. 
Without the help of an extra dimension, from e.g. QCD, we know that when the anomalous dimension $\gamma\sim1$, the theory will experience a phase transition and create bound states \cite{Fukuda:1976zb,Higashijima:1983gx,Cohen:1988sq,Miransky:1994vk}; another possibility is that the theory becomes a scale-invariant theory, which becomes the gapless unparticles discussed in \cite{Pimentel:2025rds,Yang:2025apy}. 
To understand some general features of strongly coupled dynamics in the gapped phase during inflation, we will consider unparticles living in a five-dimensional geometry, which is the product space of a four-dimensional de Sitter ($\mathrm{dS}_4$) times a circle of radius $R$. 
We generate gapped unparticles in $\mathrm{dS}_4$ by performing \textit{dimensional reduction} to the five-dimensional gapless unparticles. 
The structure is similar to Kaluza-Klein (KK) theories, but we fix the radius of the compact circle to acquire a controllable effective mass. In this scenario $M_{\rm gap}\sim 1/R$, so the radius of the circle controls the size of the gap. 
Strongly coupled spectator fields during inflation can be organized according to parameters such as: the presence of mass gap, anomalous dimensions, and the structure of couplings to the inflaton. 
Gapped unparticles realized via compactification provide the simplest setup where all of these coexist in a calculable manner. 

If the KK field theory becomes strongly coupled, we can consider its holographic dual. 
The holographic model building is widely used in Randall-Sundrum (RS) scenarios \cite{Randall:1999ee,Randall:1999vf}, Karch-Randall (KR) braneworld scenarios \cite{Karch:2000gx,Karch:2001cw,Karch:2020iit}, and many other string phenomenology models. 
In our case, the gravity dual is a crunching cosmology in a six-dimensional Anti-de Sitter (AdS) bulk, with a circle of radius changing along a ``holographic RG flow" direction. The radius of the circle can be interpreted as an additional scalar field that triggers the RG flow. 
The five-dimensional geometry we discussed above lives on the UV boundary of $\mathrm{AdS}_6$. 
This holographic construction comes from the fact that AdS spacetime has a dS slicing. 
In the usual context of holography, the dual of a thermal CFT is controlled by a competition between two geometries, the thermal AdS and AdS-Schwarzschild black hole. 
Our construction includes a CFT defined on $\mathrm{dS}_4\times S^1_R$, whose dual geometries will be a topological black hole and a bubble of nothing, as introduced in \cite{Maldacena:2012xp}. 
One can regard these geometries as performing double Wick rotation to the well-known thermal AdS and AdS-Schwarzschild black hole case. 
In our context, these geometries act more like a path to induce the boundary two-point functions. 
We will not discuss the quantum gravity aspects of these geometries, instabilities, etc. 

Our main contribution is to compute the trispectra in the collapsed limit and to offer practical, systematic constructions of strongly coupled sectors in the gapped phase. 
We emphasize that the study of strongly coupled sectors in the early universe is still at an early stage. 
Our attempt to organize this study into mass gap regimes and introduce the mass gap to unparticles is a crude approach to extract the most general features of the resulting correlators. 
Much remains to be done to embed these scenarios in a more realistic setup.

\subsection*{Outline}
The Section \ref{sec:dS_4pt} establishes the framework of gapped states construction and effective interactions in the language of EFT of inflation. 
A detailed discussion on the five-dimensional CFT two-point function we consider is provided. 
We derive the four-point function for conformally coupled scalars in de Sitter space in the collapsed limit, exchanging gapped unparticles at tree-level, using the in-in formalism. 
We analyze the phenomenological consequences in Section \ref{sec:comments_pheno}, mainly on the collapsed limit behaviors of trispectra and possible distinct signatures of extra dimensions. 
We offer a strategy to distinguish gapped unparticles from free massive scalars. 
Section \ref{sec:Outlook} concludes the key observations and provides some interesting questions for future exploration. 

The technical supplements are provided in the Appendices. 
Appendix \ref{app:von_mises} shows the essential computation of distributed localized interaction with von Mises distribution. 
In Appendix \ref{app:free_theory}, we present the dimension reduction of five-dimensional conformally coupled scalars. 
Appendix \ref{app:inf_tri} briefly introduces the weight-shifting process and contains the collapsed limit result for inflationary trispectra. 
In Appendix \ref{app:holography}, we present the derivation of the five-dimensional MFT two-point function in Section \ref{sec:dS_4pt} from six-dimensional geometries in the holographic perspective. 
Both geometries---six-dimensional topological black hole and bubble of nothing---are discussed. 

\subsection*{Main Results}
The main results of this paper are highlighted here. 
\begin{itemize}
    \item Equations \eqref{res:conformal_transformation} are the conformal transformations from five-dimensional Minkowski spacetime $\mathds{R}^{4,1}$ to five-dimensional geometry $\mathrm{dS}_4\times S^1_R$ composed from four-dimensional de Sitter spacetime and a circle. 
    These transformations can be generalize to other dimensions and are key components in the study of conformal field theory in curved spacetime. 
    \item Equation \eqref{res:strong_massive_collapsed_limit} is the analytic form for the four-point function of conformally coupled scalars in the collapsed limit, incorporating tree-level gapped unparticle mediators with generic effective masses and scaling dimensions. 
    \item Figure \ref{fig:fixed_mu_vary_Delta} shows that the scaling dimension controls the envelope of the oscillation, without affecting the frequency. 
    \item Figure \ref{fig:fixed_Delta_vary_mu} shows that the effective mass controls the frequency of the oscillatory behavior in the collapsed limit. 
    \item Figure \ref{fig:interference} shows the interference between excited modes, when the modes share a universal coupling because of localized/spread interactions. 
    \item Figure \ref{fig:brane_position} shows the effect of different positions of localized interactions. 
\end{itemize}

\subsection*{Notation and Conventions}
We will choose the metric signature to be $(-,+,+,+)$,  
and use natural units $\hbar=c=1$. 
We will use $\varphi$ for conformally coupled scalars and $\phi$ for massless fields. 
The de Sitter spatial vectors will be denoted as $\vec{x}$ or with Latin letters as $x^i$, $i=1,2,3$. 
Spacetime indices in de Sitter will be denoted using Greek letters, $\mu=0,1,2,3$. 
The five-dimensional spacetime indices will be denoted using capital Latin letters, as $x^M$, $M=0,1,2,3,4$. 
The momentum of the $n$-th external leg of a correlation function is denoted as $\vec{k}_n$. 
Its magnitude is $k_n\equiv\abs{\vec{k}_n}$. 
The sum of magnitudes of external momenta is denoted as $k_{ij}\equiv k_i+k_j$. 
The scaling dimension of an operator is denoted as $\Delta$, written in the lower index as $\mathcal{O}_\Delta$. 

\newpage
\section{de Sitter Four-Point Functions}
\label{sec:dS_4pt}

Our proxy of a strongly coupled theory, as alluded to in the introduction, will be that of a KK reduction of a 5D CFT on the manifold ${\rm dS}_4\times S^1$.\footnote{The only known explicit examples of 5D CFTs are supersymmetric, and follow from low-energy limits of string theory. For our purposes, it will suffice to consider generalized free fields with anomalous dimensions. It would of course be interesting to consider top-down examples.} The mass gap is controlled by the ratio of the radii of the circle and of dS. 
The metric is 
\begin{align}
    \dint s^2 = \frac{-\dint\eta^2+\dint \vec{x}^2}{H^2\eta^2} + \dint y^2, 
    \label{def:5D_metric}
\end{align}
and we compactify the $y$-direction with $y\equiv y+2\pi R$. 

The action we consider in this five-dimensional spacetime is 
\begin{align}
    S = S_{\mathrm{CFT}}^{\rm 5D} -\frac{1}{2} \int\dint^4x\int\dint y \sqrt{-g} \left(\nabla_M\phi\nabla^M\phi + f(y) \alpha^{-\Delta} \mathcal{O}_\Delta\nabla_M\phi\nabla^M\phi\right). 
    \label{def:5d_action}
\end{align}
Here we take the conformal field theory (CFT) to be a generalized free field (akin to mean field theory, MFT)\footnote{It would be interesting to work with interacting CFTs, in which case one needs to consider the dependence of the CFT data on e.g., the operator product expansion (OPE) coefficients. Readers can refer to \cite{Iliesiu:2018fao} for more details.}, and the CFT primary operator $\mathcal{O}_\Delta$ with \textit{scaling dimension} $\Delta$ interacts with the massless inflaton fluctuation $\phi$ through the derivatively coupled vertex, with coupling $\alpha$.  
The fluctuation of the four-dimensional zero mode of the massless field $\phi$ is related to the fluctuation of the inflaton field. 
In some literature it is denoted as $\delta\phi$. 
The profile $f(y)$ of the interaction is defined on the compactified direction, which we will discuss in the following. 

We need to perform dimensional reduction to generate the tower of gapped states in the four-dimensional effective theory. 
Since the volume of measure $\sqrt{-g}$ does not depend on $y$, it stays invariant after dimensional reduction. 
Expanding fields $\phi$ and $\mathcal{O}_\Delta$ in Fourier modes along the fifth dimension $y$ 
\begin{align}
    \phi(x^\mu,y) &= \frac{1}{\sqrt{2\pi R}} \sum_{n\in\mathds{Z}} \phi_n (x^\mu) e^{iny/R},     \label{def:phi_5d_expansion} \\
    \mathcal{O}_\Delta(x^\mu,y) &= \sum_{p\in\mathds{Z}} \mathcal{O}_{\Delta,p}(x^\mu) e^{ipy/R}, 
    \label{def:O_5d_expansion}
\end{align}
the action can be rewritten as 
\begin{align}
    \begin{aligned}
        S &= \sum_{m}S_{\mathrm{CFT},m}^{4d} + S_{\mathrm{scalar}} + S_{\mathrm{int}}, \\
        S_{\mathrm{scalar}} &= -\frac{1}{2} \sum_n \int\dint^4x \sqrt{-g} \left(\nabla_\mu\phi_n\nabla^\mu\phi_n+\frac{n^2}{R^2}|\phi_n|^2\right), \\
        S_{\mathrm{int}} &= -\frac{1}{2}\int\dint^4x\int\dint y \sqrt{-g} f(y) \alpha^{-\Delta} \mathcal{O}_\Delta\nabla_M\phi\nabla^M\phi. 
    \end{aligned}
\end{align}
To find the effective interactions in four dimension, we expand the interacting part $S_{\mathrm{int}}$ of action. 
First, we have 
\begin{align}
    &\begin{aligned}
        \nabla_M\phi\nabla^M\phi &= \nabla_\mu\phi\nabla^\mu\phi + \partial_y\phi\partial^y\phi \\
        &= \frac{1}{2\pi R} \sum_{n_1,n_2} \left(\nabla_\mu\phi_{n_1}\nabla^\mu\phi_{n_2}-\frac{n_1n_2}{R^2}\phi_{n_1}\phi_{n_2}\right) e^{i(n_1+n_2)y/R} 
    \end{aligned} \\
    &\begin{aligned}
        \Rightarrow S_{\mathrm{int}} &= -\frac{1}{2}\int\dint^4x \sqrt{-g} \frac{\alpha^{-\Delta}}{2\pi R} \sum_{p,n_1,n_2} \mathcal{O}_{\Delta,p} \left(\nabla_\mu\phi_{n_1}\nabla^\mu\phi_{n_2}-\frac{n_1n_2}{R^2}\phi_{n_1}\phi_{n_2}\right) \\
        &\times \int\dint y\ f(y) e^{i(p+n_1+n_2)y/R}. 
    \end{aligned} 
\end{align}
The inflaton perturbations are to good approximation massless, so we focus on the $m=0$ mode of $\phi$. This corresponds to $n_1=n_2=0$. 

We can have several choices for the profile $f(y)$: 
\begin{itemize}
    \item[1.] $f(y)=1$. This is the plain distribution. 
    Due to the momentum conservation on the circle, only the zero mode of $\mathcal{O}_{\Delta,p}$ can contribute to the interaction vertices, if both $\phi_n$'s are $\phi_0$. 
    The $f_{\mathrm{NL}}$ signal will be similar to the standard Cosmological Collider and there will be no interference between modes. 
    \item[2.] $f(y)=2\pi R\delta(y-y_0)$. This corresponds to an interaction localized at $y=y_0$ along the circle.
    Such localized couplings are motivated by brane-world scenarios, for example, RS models, though a concrete realization in the present context would require additional model building beyond the scope of this work.
    We fix the coordinate origin by placing the observable universe on a $\rm dS_4$-brane at $y=0$, exploiting the translational symmetry of $S^1$.\footnote{In this work, we do not attempt to stabilize the relative position of the two branes; $y_0$ is treated as a free parameter characterizing the geometric separation between the observer's brane and the interaction locus.}
    The interaction is then localized at a relative separation $y_0$ from this observer's brane, as shown in figure~\ref{fig:illus_localized_interaction}.
    \begin{figure}[t]
        \centering
        \includegraphics[width=0.7\linewidth]{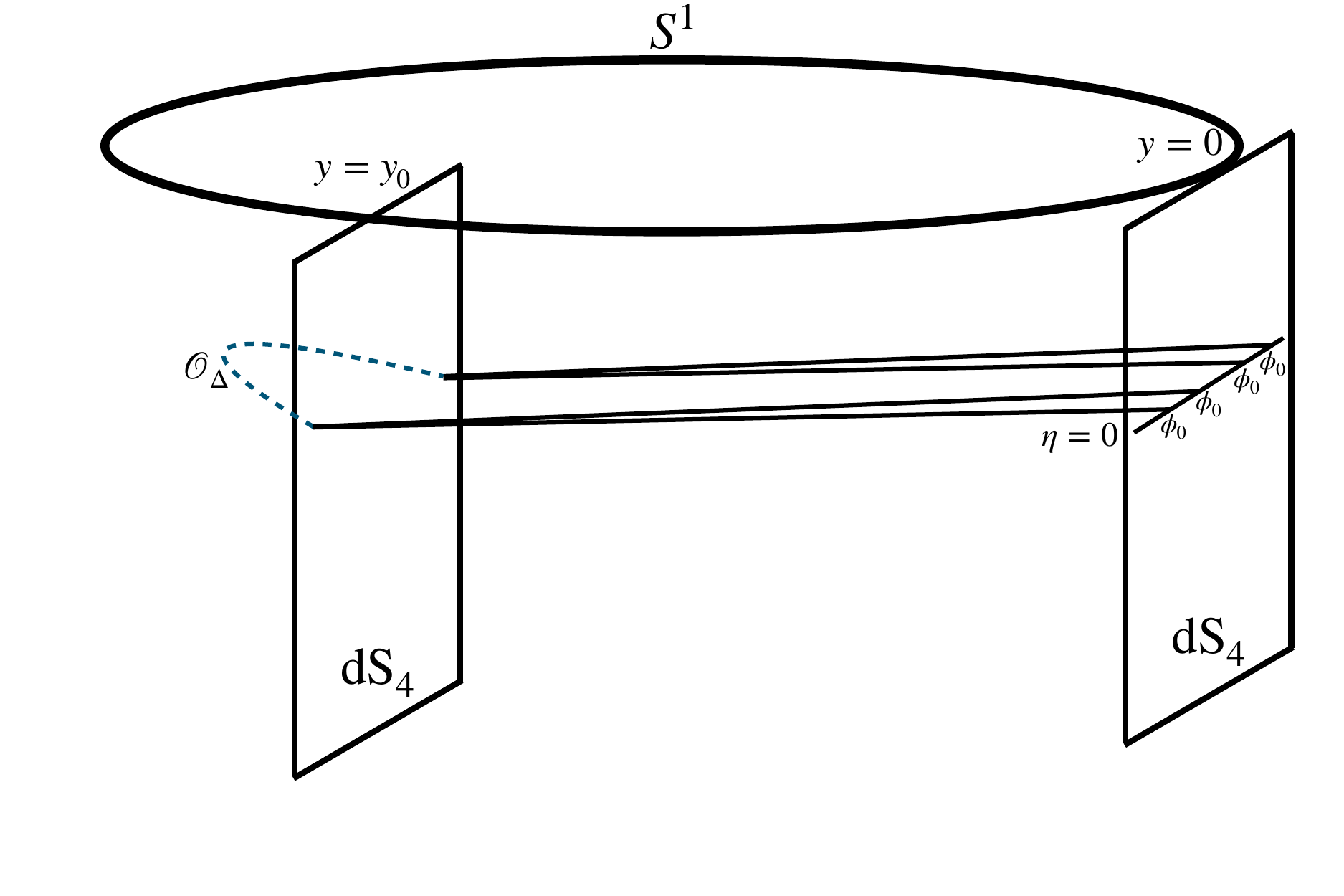}
        \caption{Illustration of the localized interaction in ${\rm dS}_4\times S^1$}
        \label{fig:illus_localized_interaction}
    \end{figure}
    Although the inflaton zero mode $\phi_0$ is $y$-independent and takes the same value everywhere on the circle, the higher-dimensional fields that mediate the interaction propagate through the bulk from $y=y_0$ to the observer at $y=0$.
    As a result, the inflaton can couple to operators localized at $y=y_0$ through their KK decomposition.
    While the underlying geometry is translationally invariant, the relative separation $y_0$ is physical because it is measured with respect to the observer's brane at $y=0$; the phase factor $e^{ipy_0/R}$ acquired by the KK modes reflects their bulk propagation from the interaction locus to the observation point.
    Performing the $y$-integration yields this factor, and the effective interaction measured on the $y=0$ brane becomes 
    \begin{align}
        S_{\mathrm{int}} &= -\frac{1}{2}\int\dint^4x \sqrt{-g} \alpha^{-\Delta} \sum_{p} e^{ipy_0/R}  \mathcal{O}_{\Delta,p} \nabla_\mu\phi_{0}\nabla^\mu\phi_{0}. 
        \label{res:5d_brane_interaction}
    \end{align}
    \item[3.] $f(y;\kappa)=\exp\left(\kappa\cos(y/R-y_0/R)\right)/(2\pi I_{0}(\kappa))$.
    This profile describes a smooth localization around $y=y_0$, given by the von Mises distribution\footnote{Our results are robust under smooth deformations of the localization profile. The von Mises distribution is chosen for concreteness and computational clarity.}, where $I_{\nu}(z)$ is the modified Bessel function of the first kind.
    This can be viewed as a thick brane centered at $y=y_0$, at a finite relative separation from the observer's brane at $y=0$.
    As in the previous case, the interaction is mediated by bulk modes that extend over the circle, so the inflaton zero mode can still couple to this localized structure.
    With the definition of the von Mises distribution reviewed in Appendix \ref{app:von_mises}, the induced effective interaction is
    \begin{align}
        S_{\mathrm{int}} &= -\frac{1}{2}\int\dint^4x \sqrt{-g} \frac{\alpha^{-\Delta}} {2\pi}\sum_{p} e^{ipy_0/R} \frac{I_{p}(\kappa)}{I_0(\kappa)} \mathcal{O}_{\Delta,p} \nabla_\mu\phi_{0}\nabla^\mu\phi_{0}. 
        \label{res:5d_fat_brane_interaction}
    \end{align}
\end{itemize}

From now on, we will simply denote $\phi_0\equiv\phi$. 
To compute the four-point functions $\langle\phi_{\vec{k}_1}\phi_{\vec{k}_2}\phi_{\vec{k}_3}\phi_{\vec{k}_4}\rangle$, we can write simplified interactions based on \eqref{res:5d_brane_interaction} and \eqref{res:5d_fat_brane_interaction}, and perform ``weight-shifting" \cite{Arkani-Hamed:2018kmz,Baumann:2019oyu} to generate the results for these target interactions. 
We will first consider the interaction between gapped unparticles $\mathcal{O}_{\Delta,p}$ and conformally coupled scalars $\varphi$, as 
\begin{align}
    &S = \sum_{p}S_{\mathrm{CFT},p}^{4d} - \frac{1}{2} \int\dint^4 x\sqrt{-g} \left(\nabla_\mu\varphi\nabla^\mu\varphi+2H^2\varphi^2+\sum_{p}\alpha_p^{2-\Delta} \mathcal{O}_{\Delta,p}\varphi^2\right), 
    \label{def:4d_effective_action} \\
    &\alpha_p^{2-\Delta} \equiv \alpha^{2-\Delta} e^{ipy_0/R} \quad \mathrm{localized}, \quad
    \alpha_p^{2-\Delta} \equiv \alpha^{2-\Delta} \frac{e^{ipy_0/R}}{2\pi}\frac{I_{p}(\kappa)}{I_0(\kappa)} \quad \mathrm{spread}, \label{def:cc_coupling}
\end{align}
where $\alpha_p$ is the coupling constant. 
The mode function in Bunch-Davies vacuum for conformally coupled scalars is 
\begin{align}
    \hat{f}_{\varphi}(k,\eta) = (-H\eta)\frac{e^{-ik\eta}}{\sqrt{2k}}. 
\end{align}
The Green's functions with and without time-ordering can be defined from the mode functions: 
\begin{align}
    \begin{aligned}
            &G_{++}(k;\eta_1,\eta_2) = \hat{f}(k,\eta_1) \hat{f}^*(k,\eta_2) \theta(\eta_1-\eta_2) + \hat{f}^*(k,\eta_1) \hat{f}(k,\eta_2) \theta(\eta_2-\eta_1), \\
            &G_{+-}(k;\eta_1,\eta_2) = \hat{f}^*(k,\eta_1) \hat{f}(k,\eta_2), \\
            &G_{-+}(k;\eta_1,\eta_2) = G_{+-}^*(k;\eta_1,\eta_2), \quad G_{--}(k;\eta_1,\eta_2) = G_{++}^*(k;\eta_1,\eta_2). 
    \end{aligned}
\end{align}

Since we only consider MFTs, the two-point functions of $\mathcal{O}_{\Delta,p}$ will not mix between different KK levels. 
The two-point function of a scalar CFT in flat space in Lorentzian signature (after Wick rotation) is completely fixed by conformal symmetries, which is 
\begin{equation}
    \langle\mathcal{O}_\Delta(t_1,\vec{a}_1)\mathcal{O}_\Delta(t_2,\vec{a}_2)\rangle_{\mathrm{Mink}} = \frac{1}{\Big(-(t_1 - t_2)^2 + (\vec{a}_1-\vec{a}_2)^2\Big)^\Delta}. 
\end{equation}
The unitarity bound for scalar CFT two-point functions in $D$ dimensions is $\Delta\geq(D-2)/2$. 
In five dimensions, unitarity requires the scaling dimension to be $\Delta\geq3/2$, where the critical value $\Delta=3/2$ corresponds to conformally coupled scalars. 
An introduction on conformally coupled scalars in more detail is presented in Appendix \ref{app:free_theory}. 
It is worth noticing that, the critical value $\Delta=3/2$ is a benchmark for our results, because it corresponds to both conformal and free massive scalar fields. 

The conformal transformation relating a five-dimensional Minkowski spacetime $\mathds{R}^{4,1}$ to $\mathrm{dS}_4 \times \mathds{R}$ reads 
\begin{align}
    \begin{aligned}
        a_1 &= \frac{\eta\sinh(H y)}{H^2\eta^2+2H\eta\cosh(H y)-H^2\vec{x}^2+1}\\
        a_{i+1} &= \frac{x_i}{H^2\eta^2+2H\eta\cosh(H y)-H^2\vec{x}^2+1}\\
        t&= \frac{(-H\eta^2+H\vec{x}^2+1/H)/2}{H^2\eta^2+2H\eta\cosh(H y)-H^2\vec{x}^2+1}. 
    \end{aligned} \label{res:conformal_transformation}
\end{align}
Therefore, the metrics are related by a Weyl transformation 
\begin{align}
    \begin{aligned}
    \dint s^2&=-\dint t^2+\dint\vec{a}^2=\frac{H^2\eta^2 }{\left(H^2\eta^2+2H\eta\cosh(H y)-H^2\vec{x}^2+1\right)^2} \left(\frac{-\dint\eta^2+\dint\vec{x}^2}{H^2\eta^2}+\dint y^2 \right)\\
    &\equiv\Omega(Y)^2 \left(\frac{-\dint\eta^2+\dint\vec{x}^2}{H^2\eta^2}+\dint y^2 \right), 
    \end{aligned}
\end{align}
where $\vec{a}$ is four-dimensional. 
The two-point function of our CFT in $\mathrm{dS}_4 \times \mathds{R}$ can be derived as 
\begin{align}
    g(Y_1,Y_2) \equiv \langle \mathcal{O}_\Delta(\eta_1,\vec{x}_1,y_1) \mathcal{O}_\Delta(\eta_2,\vec{x}_2,y_2) \rangle = \Omega(Y_1)^\Delta\Omega(Y_2)^\Delta \langle \mathcal{O}_\Delta(t_1,\vec{a}_1) \mathcal{O}_\Delta(t_2,\vec{a}_2) \rangle. 
\end{align}
Now we compactify the extra dimension $\mathds{R}$ into a circle with finite radius $R$ with periodic boundary condition. 
We need to sum up the windings around $S^1_R$ to obtain the full two-point function: 
\begin{align}
    g(Y_1,Y_2) = \sum_{m=-\infty}^{\infty} g_m(Y_1,Y_2) = \sum_{m=-\infty}^{\infty} g_0(\eta_1,\eta_2;\abs{\vec{x}_1-\vec{x}_2};y_1-y_2+2\pi mR). 
\end{align}
Denoting $\beta\equiv2\pi R$, the five-dimensional two-point function is 
\begin{align}
    \langle\mathcal{O}_\Delta(\eta_1,\vec{x},y) \mathcal{O}_\Delta(\eta_2,\vec{0},0)\rangle = \sum_{m=-\infty}^{\infty} \frac{(H^2\eta_1\eta_2)^\Delta}{\left(-\eta_1^2-\eta_2^2+\vec{x}^2+2\eta_1 \eta_2  \cosh(H(y+m\beta))\right)^\Delta}. 
    \label{res:5d_2pt_gapped_cft}
\end{align}
To perform the dimensional reduction, we need to find the Fourier coefficients of $y$-direction. 
With the Fourier index $p$, which will label the effective mass of the gapped unparticles, we can take the energy cutoff and only consider the excitation of states from $-N$ to $N$. 
The two-point function can be written in the Schwinger parametrization: 
\begin{align}
    &\langle\mathcal{O}_{\Delta}(\eta_1,\vec{x},y) \mathcal{O}_{\Delta}(\eta_2,\vec{0},0)\rangle \nonumber \\
    =& \sum_{m=-\infty}^{\infty} \frac{\left(H^2\eta_1\eta_2\right)^\Delta}{\left(-\eta_1^2-\eta_2^2+\vec{x}^2+2\eta_1 \eta_2  \cosh(H(y+m\beta))\right)^\Delta} \\
    =& \left(H^2\eta_1\eta_2\right)^\Delta \frac{i^{-\Delta}}{\Gamma(\Delta)} \sum_{m=-\infty}^{\infty} \int_0^\infty \dint s\ s^{\Delta-1} e^{is\left(-\eta_1^2-\eta_2^2+\vec{x}^2+2\eta_1 \eta_2 \cosh(H(y+m\beta))\right)}. 
\end{align}

The Fourier transform of $\sum\exp(2is\eta_1\eta_2\cosh(H(y+m\beta)))$ is 
\begin{align}
    g(p) &= \frac{1}{H\beta}\int_{0}^{H\beta} \dint(H y)\ \left(\sum_{m=-\infty}^\infty e^{2is\eta_1\eta_2\cosh(H(y+m\beta))}\right) e^{-i2\pi\frac{p}{H\beta}H y} \nonumber \\
    &= \frac{1}{H\beta} \int_{-\infty}^{\infty} \dint (H y)\ e^{2is\eta_1\eta_2\cosh(H y)} e^{-i2\pi\frac{p}{H\beta}H y} \nonumber \\
    &= \frac{1}{H\pi R} K_{i\frac{p}{HR}} \left(-2is\eta_1\eta_2\right), \quad p\in\mathds{Z}, 
\end{align}
where $K_{\nu}(z)$ is the modified Bessel function of the second kind. 
For simplicity, and also for comparison, we define the four-dimensional effective mass of unparticle KK modes $p/(HR)$ as $\mu$.  
It is of great interest to inspect the small $\eta_1$, $\eta_2$ limit, because it corresponds to the collapsed limit of the trispectra. 
The limiting form of Bessel K function gives us 
\begin{align}
    &K_{\nu}(z)\Big|_{z\rightarrow0} \sim \frac{1}{2} \Gamma(-\nu)\left(\frac{1}{2} z\right)^{\nu} + \frac{1}{2} \Gamma(\nu)\left(\frac{1}{2} z\right)^{-\nu} \ \mathrm{when   }\ \nu\notin\mathds{Z}, \\
    \Rightarrow& K_{i\mu} (-2is\eta_1\eta_2) \sim \frac{1}{2} \Gamma\left(-i\mu\right) e^{\pi\mu/2} \left(s\eta_1\eta_2\right)^{i\mu} + \frac{1}{2} \Gamma\left(i\mu\right) e^{-\pi\mu/2} \left(s\eta_1\eta_2\right)^{-i\mu}. 
\end{align}
After performing the $s$-integral, the four-dimensional two-point function for individual $\mu$'s in the small $\eta_1$, $\eta_2$ limit becomes 
\begin{align}
    \langle\mathcal{O}_\Delta(\eta_1,\vec{x})\mathcal{O}_\Delta(\eta_2,\vec{0})\rangle_\mu &= \frac{H^{2\Delta-1}}{2\pi R} \frac{1}{\Gamma(\Delta)} \Bigg(\left(\frac{\eta_1 \eta_2}{\vec{x}^2}\right)^{\Delta+i\mu} \Gamma\left(-i\mu\right)\Gamma\left(\Delta+i\mu\right) \nonumber \\
    &+ \left(\frac{\eta_1 \eta_2}{\vec{x}^2}\right)^{\Delta-i\mu} \Gamma\left(i\mu\right) \Gamma\left(\Delta-i\mu\right)\Bigg) + \dots
\end{align}
Using the three-dimensional spatial Fourier transform  
\begin{align}
    \int\dint^{3}x\ e^{i\vec{k}\cdot\vec{x}}\abs{x}^{-2a} = 8\pi^{3/2} 2^{-2a} k^{2a-3} \frac{\Gamma\left(\frac{3}{2}-a\right)}{\Gamma(a)}, 
\end{align}
we get the two-point function $\langle\mathcal{O}_{\Delta}(\eta_1,k)\mathcal{O}_{\Delta}(\eta_2,k)\rangle_\mu$ in momentum space in small $k$ limit: 
\begin{align}
    \langle\mathcal{O}_{\Delta}(\eta_1,k)\mathcal{O}_{\Delta}(\eta_2,k)\rangle_\mu &= \frac{4\sqrt{\pi}H^{2\Delta-1}}{\Gamma(\Delta)k^3R} \Bigg(\left(\frac{k^2\eta_1\eta_2}{4}\right)^{\Delta+i\mu} \Gamma\left(-i\mu\right) \Gamma\left(\frac{3}{2}-\Delta-i\mu\right) \nonumber \\
    &+\left(\frac{k^2\eta_1\eta_2}{4}\right)^{\Delta-i\mu} \Gamma\left(i\mu\right) \Gamma\left(\frac{3}{2}-\Delta+i\mu\right)\Bigg) + \dots
\end{align}
As a sanity check, notice that when $\Delta=3/2$, we recover the known result of free massive scalar exchanges \cite{Arkani-Hamed:2015bza,Arkani-Hamed:2018kmz}. The factor of $\Delta$ acts as an effective spatial dimension for the particle, modifying how it redshifts with time.

With all the ingredients at hand, we can compute the correlator $\langle\varphi_{\vec{k}_1}\varphi_{\vec{k}_2}\varphi_{\vec{k}_3}\varphi_{\vec{k}_4}\rangle$ using the ``in-in" formalism. 
The correlation function with two-site tree-level exchange can always be divided into four sectors \cite{Schwinger:1960qe,Keldysh:1964ud,Jordan:1986ug,Calzetta:1986ey,Maldacena:2002vr,Chen:2010xka,Arkani-Hamed:2015bza}: 
\begin{equation}
    \begin{aligned}
        \langle\varphi_{\vec{k}_1}\varphi_{\vec{k}_2}\varphi_{\vec{k}_3}\varphi_{\vec{k}_4}\rangle &\equiv (2\pi)^3\delta^{(3)} \left(\sum_i\vec{k}_i\right) \langle\varphi_{\vec{k}_1}\varphi_{\vec{k}_2}\varphi_{\vec{k}_3}\varphi_{\vec{k}_4}\rangle', \\
        \langle\varphi_{\vec{k}_1}\varphi_{\vec{k}_2}\varphi_{\vec{k}_3}\varphi_{\vec{k}_4}\rangle' &\equiv I_{++} + I_{+-} + I_{-+} + I_{--}. 
        \label{def:dS_4pt}
    \end{aligned}
\end{equation}
Since we are working at tree-level, the momentum magnitude $k$ of the exchanged gapped unparticle is $k\equiv|\vec{k}_1+\vec{k}_2|$ in the $s$-channel. 
Diagrammatically, $\langle\varphi_{\vec{k}_1}\varphi_{\vec{k}_2}\varphi_{\vec{k}_3}\varphi_{\vec{k}_4}\rangle'$ can be expressed as 
\begin{equation}
    \langle\varphi_{\vec{k}_1}\varphi_{\vec{k}_2}\varphi_{\vec{k}_3}\varphi_{\vec{k}_4}\rangle' \equiv \raisebox{-0.7cm}{\includegraphics[width=2.5cm]{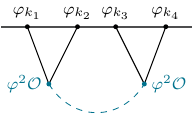}} + \raisebox{-0.7cm}{\includegraphics[width=2.5cm]{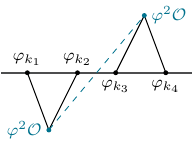}} + \raisebox{-0.7cm}{\includegraphics[width=2.5cm]{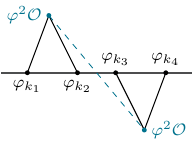}} + \raisebox{-0.3cm}{\includegraphics[width=2.5cm]{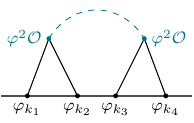}}.
\end{equation}
The full expression of the four-point function requires a sum over permutations. 

The integral $I_{\pm\pm}$ in the collapsed limit gives 
\begin{align}
    \begin{aligned}
        &I_{\pm\pm,\mu} = \frac{2^2\alpha_p^{4-2\Delta}H^{2\Delta-1}\eta_0^4}{16k_1k_2k_3k_4} \frac{4\sqrt{\pi}}{\Gamma(\Delta)k^3R} (\mp i)(\mp i) \int_{-\infty}^{0} \int_{-\infty}^{0} \frac{\dint\eta_1}{\eta_1^2}\frac{\dint\eta_2}{\eta_2^2}e^{\pm ik_{12}\eta_1} e^{\pm ik_{34}\eta_2} \\
        \times& \Bigg(\left(\frac{k^2\eta_1\eta_2}{4}\right)^{\Delta+i\mu} \Gamma\left(-i\mu\right) \Gamma\left(\frac{3}{2}-\Delta-i\mu\right) + \left(\frac{k^2\eta_1\eta_2}{4}\right)^{\Delta-i\mu} \Gamma\left(i\mu\right) \Gamma\left(\frac{3}{2}-\Delta+i\mu\right)\Bigg), 
    \end{aligned}
\end{align}
where $\eta_0$ is the late-time cutoff for the bulk-to-boundary propagator of $\varphi$ and $2^2$ is a symmetry factor. 
The appearance of this late-time cutoff indicates that the correlation functions of conformally coupled scalars $\varphi$ are redshifted away when $\eta_0\rightarrow0$. We only keep track of the leading behavior at late times, the amplitude of the ``growing mode."  From these correlation functions of $\varphi$, we can later derive the correlation functions of $\phi$. 
Performing the $\eta$-integrals: 
\begin{align}
    J_{\pm}=\int_{-\infty}^0 \dint \eta_1\ e^{\pm ik_{12}\eta_1} \eta_1^{\Delta-2+i\mu} = (\pm i)(\pm i)^{\Delta+i\mu} k_{12}^{1-\Delta-i\mu} \Gamma(\Delta-1+i\mu). 
\end{align}

Defining the ratio of momenta $u\equiv k/k_{12}$ and $v\equiv k/k_{34}$ to keep the notation consistent with \cite{Arkani-Hamed:2018kmz,Pimentel:2025rds}, the final collapsed-limit result is 
\begin{align}
    \begin{aligned}
        &\langle\varphi_{\vec{k}_1}\varphi_{\vec{k}_2}\varphi_{\vec{k}_3}\varphi_{\vec{k}_4}\rangle'_\mu = \frac{\eta_0^4\alpha_p^{4-2\Delta}H^{2\Delta-1}}{16k_1k_2k_3k_4} \frac{2^3\sqrt{\pi}}{\Gamma(\Delta)kR} \Bigg(\Big(1+\cos\big(\pi(\Delta+i\mu)\big)\Big) \left(\frac{uv}{4}\right)^{\Delta-1+i\mu} \\
        &\times \Gamma\left(-i\mu\right) \Gamma\left(3/2-\Delta-i\mu\right) \Gamma(\Delta-1+i\mu)^2 + \Big(1+\cos\big(\pi(\Delta-i\mu)\big)\Big) \left(\frac{uv}{4}\right)^{\Delta-1-i\mu} \\
        &\times \Gamma\left(i\mu\right) \Gamma\left(3/2-\Delta+i\mu\right) \Gamma(\Delta-1-i\mu)^2\Bigg). 
        \label{res:strong_massive_collapsed_limit}
    \end{aligned}
\end{align}
If we take $\Delta=3/2$, the four-point function becomes 
\begin{align}
    \begin{aligned}
        &\langle\varphi_{\vec{k}_1}\varphi_{\vec{k}_2}\varphi_{\vec{k}_3}\varphi_{\vec{k}_4}\rangle'_\mu\left[\Delta=\frac{3}{2}\right] \\
        =& \frac{\eta_0^4\alpha_pH^{2}}{16k_1k_2k_3k_4} \frac{16\sqrt{2}}{kR} \Bigg((1+i\sinh\pi\mu)\left(\frac{uv}{4}\right)^{\frac{1}{2}+i\mu} \Gamma\left(-i\mu\right)^2 \Gamma\left(\frac{1}{2}+i\mu\right)^2 \\
        +& (1-i\sinh\pi\mu)\left(\frac{uv}{4}\right)^{\frac{1}{2}-i\mu} \Gamma\left(i\mu\right)^2 \Gamma\left(\frac{1}{2}-i\mu\right)^2\Bigg), 
    \end{aligned}
\end{align}
which is the well-known result for free scalar exchanges \cite{Arkani-Hamed:2015bza,Arkani-Hamed:2018kmz}. 
Note that \eqref{res:strong_massive_collapsed_limit} decays as $e^{-\pi\mu}$ for large $\mu$. 

For future simplicity, we define the essential part that only depends on the dimensionless variables $u$ and $v$ to be the \textit{master function} $F^{(\mu)}(u,v)$: 
\begin{align}
    &F^{(\mu)}(u,v) \equiv \Big(1+\cos\big(\pi(\Delta+i\mu)\big)\Big) \left(\frac{uv}{4}\right)^{\Delta-1+i\mu} \Gamma\left(-i\mu\right) \Gamma\left(3/2-\Delta-i\mu\right) \Gamma(\Delta-1+i\mu)^2 \nonumber \\
    &+ \Big(1+\cos\big(\pi(\Delta-i\mu)\big)\Big) \left(\frac{uv}{4}\right)^{\Delta-1-i\mu} \Gamma\left(i\mu\right) \Gamma\left(3/2-\Delta+i\mu\right) \Gamma(\Delta-1-i\mu)^2, 
    \label{res:master_function}
\end{align}
where $\mu$ denotes the effective mass. 
If we want to see the significant interference between several modes, the effective mass $\mu$ needs to be $\mu\sim1/\pi$, in order to make the amplitude between modes comparable. 
A detailed discussion on phenomenology will be presented in Section \ref{sec:comments_pheno}.  

The method we used above to expand the two-point function of gapped unparticles is suitable for computing the collapsed limit of trispectra. In this limit, the exchange momentum is very small, i.e. $u,v\ll 1$. The expansion we constructed is therefore tailored to probe this regime. 

A non-zero bispectrum is expected from diagrams involving one quadratic and one linear mixing interaction. In order to properly compute it we need different kinematics, namely $u\ll 1$ with $v\to1$, or vice versa. This regime lies far from the domain where we have analytic control.
It would nonetheless be very interesting to investigate the potential bispectrum, either numerically or with other analytic techniques. Our focus instead remains on highlighting potentially distinctive signatures in non-Gaussianity.
\section{Comments on Phenomenology}
\label{sec:comments_pheno}

In this section, we will compare the curves of master functions with various parameters. 
Notice that these are the plots for conformally coupled scalar correlation functions. For the inflaton, the plots are qualitatively similar, as presented in Appendix \ref{app:inf_tri}.
Particularly, when the effective gap $\mu$ is $\sim O(1/\pi)$, it is possible to observe interference between excited states, as mentioned in Section \ref{sec:dS_4pt}. 
This peculiar behavior is the direct consequence of the localized interactions \eqref{res:5d_brane_interaction} and \eqref{res:5d_fat_brane_interaction}. 

\begin{figure}[h]
    \centering
    \includegraphics[width=0.6\linewidth]{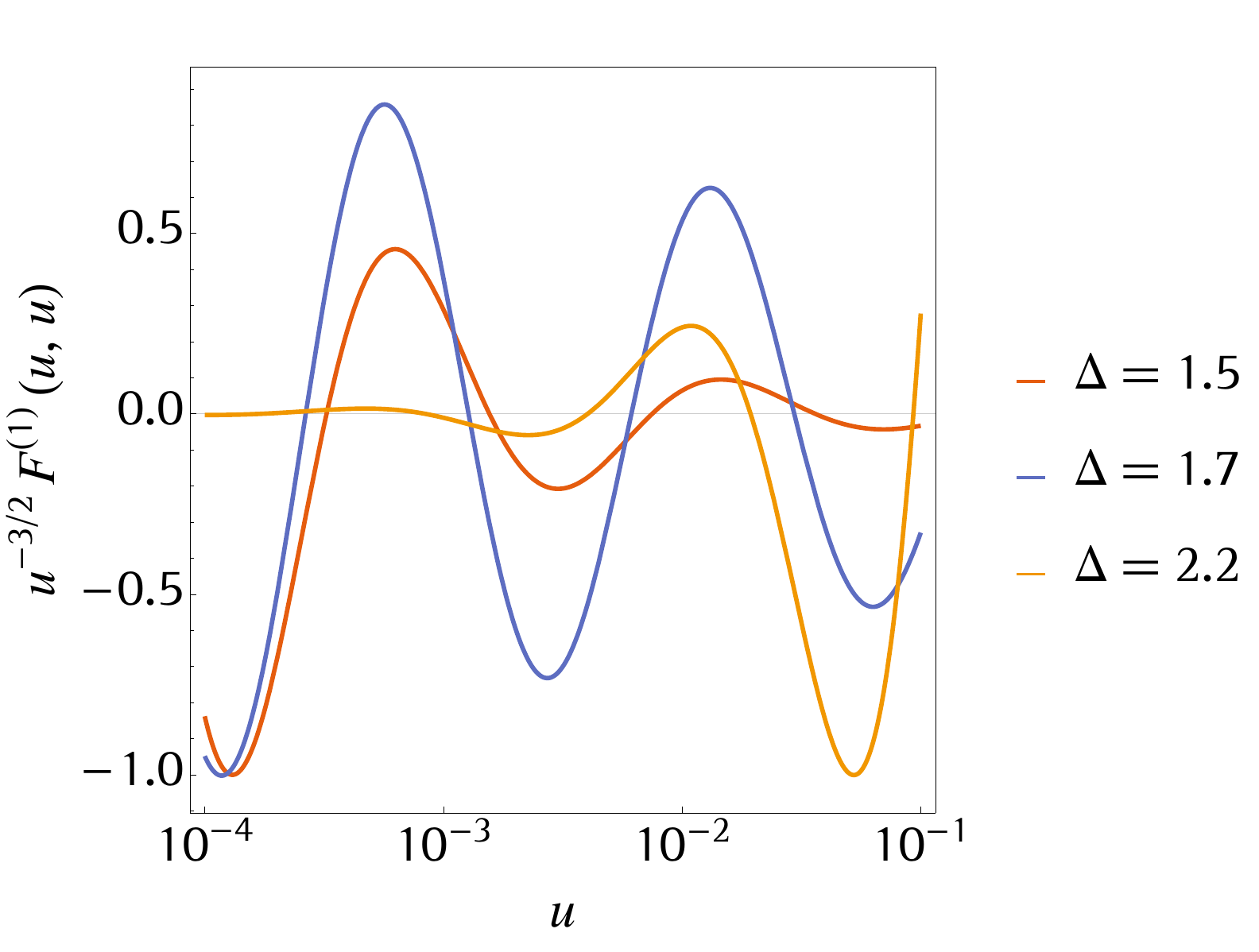}
    \caption{Behaviors in the collapsed limit when we fix $\mu\equiv1$, as examples. In order to visually enhance the magnitudes, we rescaled $F^{(1)}(u,u)$ by $u^{-\Delta}$ and rescaled the minima to $-1$. We can see that when we fix the effective mass, the scaling dimension controls the envelope of the shapes.}
    \label{fig:fixed_mu_vary_Delta}
\end{figure}

Figure \ref{fig:fixed_mu_vary_Delta} shows the behaviors of modes of identical effective mass $\mu=1$. 
We set $u=v$ to reduce the variables. 
We rescaled the master function $F^{(\mu)}(u,v)$ by $u^{-3/2}$, and rescaled the minima to $-1$ for shape comparison. 
We can see that the envelopes of the oscillations depend on $\Delta$, which can be read from \eqref{res:master_function}. 
It is worth reminding that $\Delta=1.5$ corresponds to all free massive scalar exchanges. 
The real part in the power of $u$ for $F^{(\mu)}(u,u)$ is $u^{2\Delta-2}$. 
The frequency of oscillations is controlled by $\mu$, as shown in \eqref{res:master_function} that only $\mu$ contributes to the imaginary part of the power of $u$ and $v$. 
We can interpret this change of envelop as an effect of strong quantum corrections. 
At weak coupling \cite{Sloth:2006az,Kumar:2009ge,Krotov:2010ma,Marolf:2010zp,Marolf:2010nz,Marolf:2011sh,Jatkar:2011ju,Marolf:2012kh,Rajaraman:2015dta,Rajaraman:2016nvv,Chen:2016nrs,Lu:2021wxu,DiPietro:2021sjt} it has been known for a while that when we consider loop corrections to tree-level correlators, we tend to increase the dilution, which means the correlators decay faster in the collapsed limit. 
The same loop corrections give rise to the large anomalous dimensions for strongly coupled unparticles. 

\begin{figure}[h]
    \centering
    \begin{subfigure}{0.32\textwidth}
        \includegraphics[width=\textwidth]{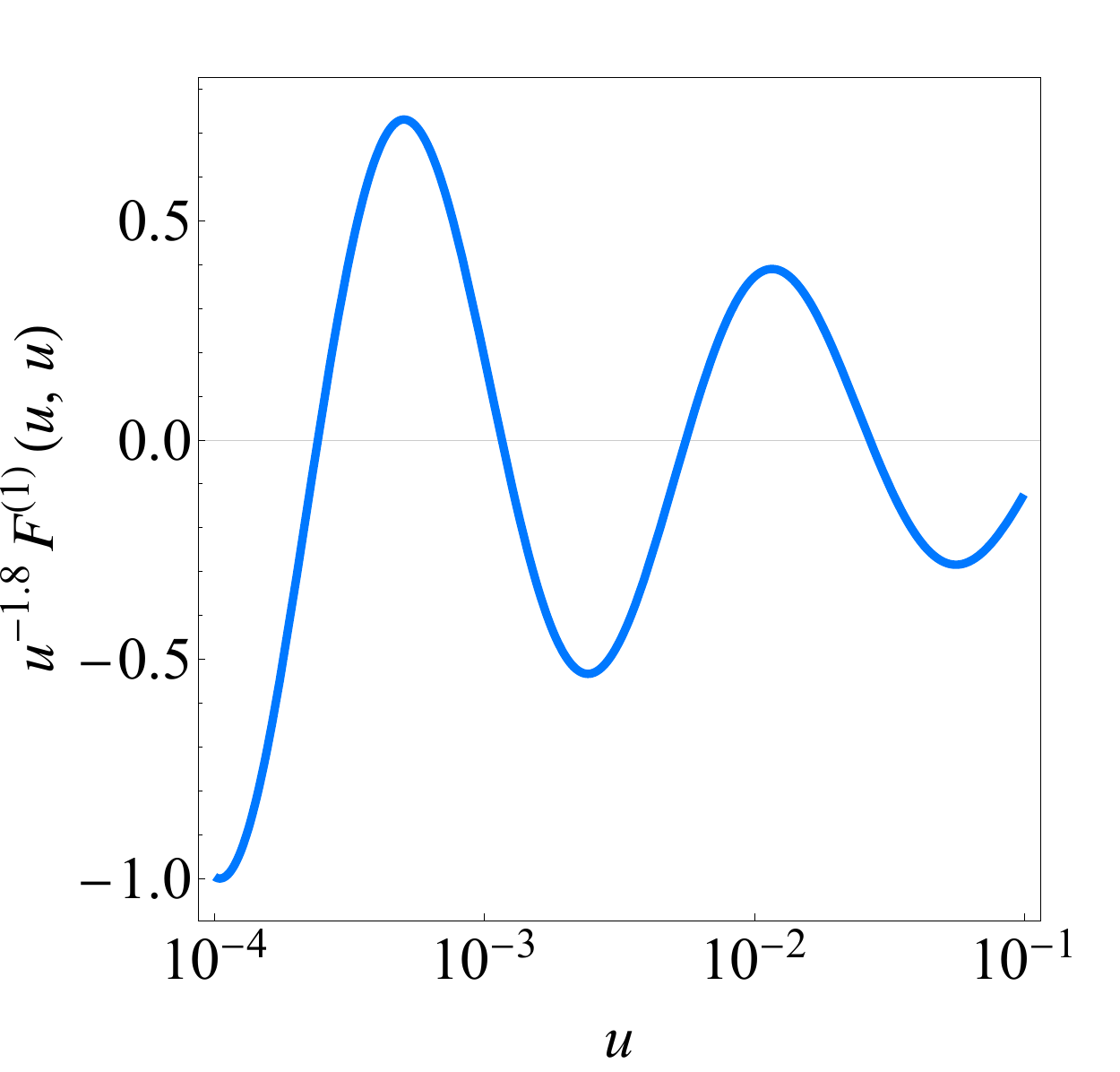}
        \caption{$\mu=1$}
    \end{subfigure}
    \hfill
    \begin{subfigure}{0.32\textwidth}
        \includegraphics[width=\textwidth]{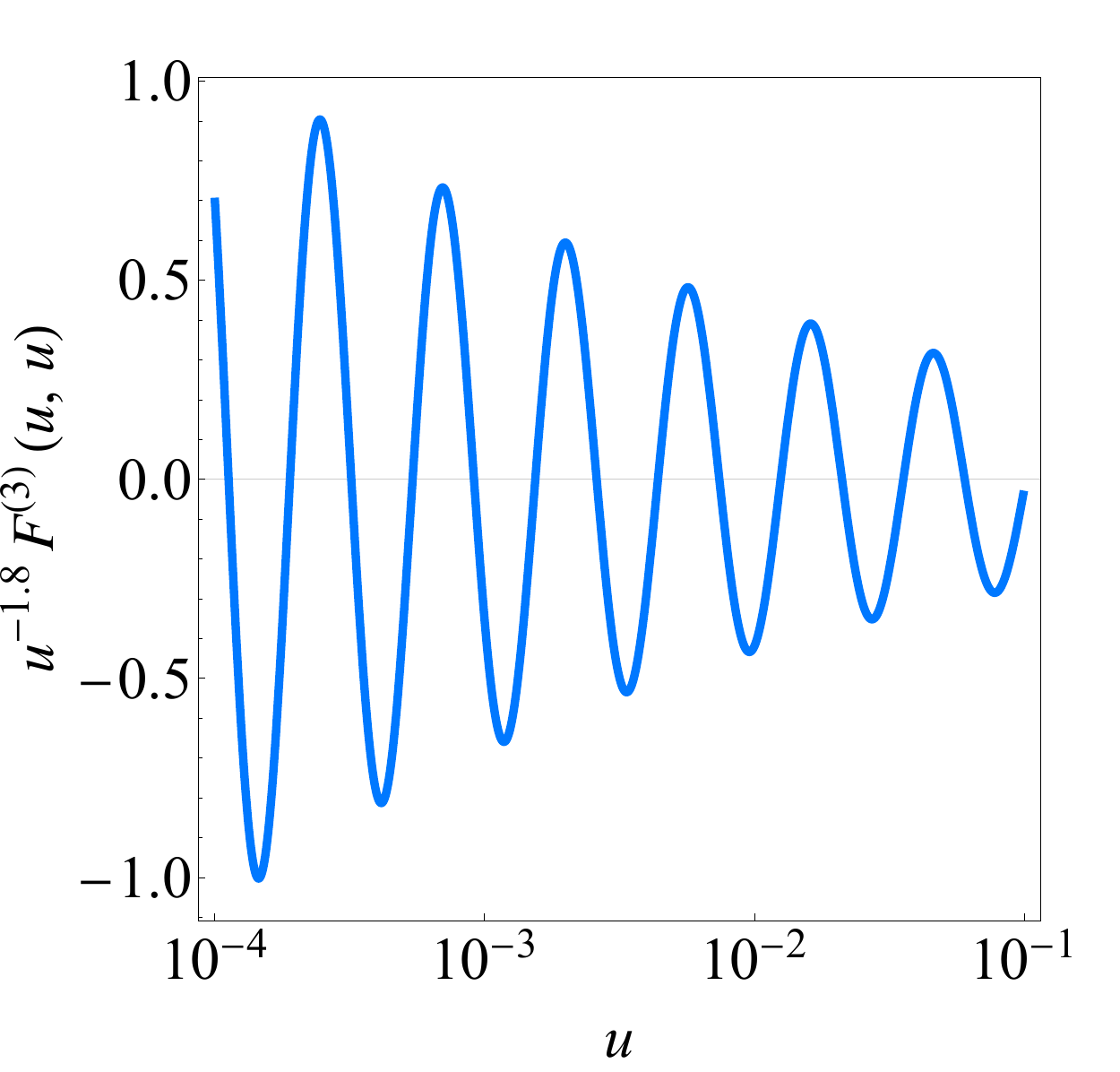}
        \caption{$\mu=3$}
    \end{subfigure}
    \hfill
    \begin{subfigure}{0.32\textwidth}
        \includegraphics[width=\textwidth]{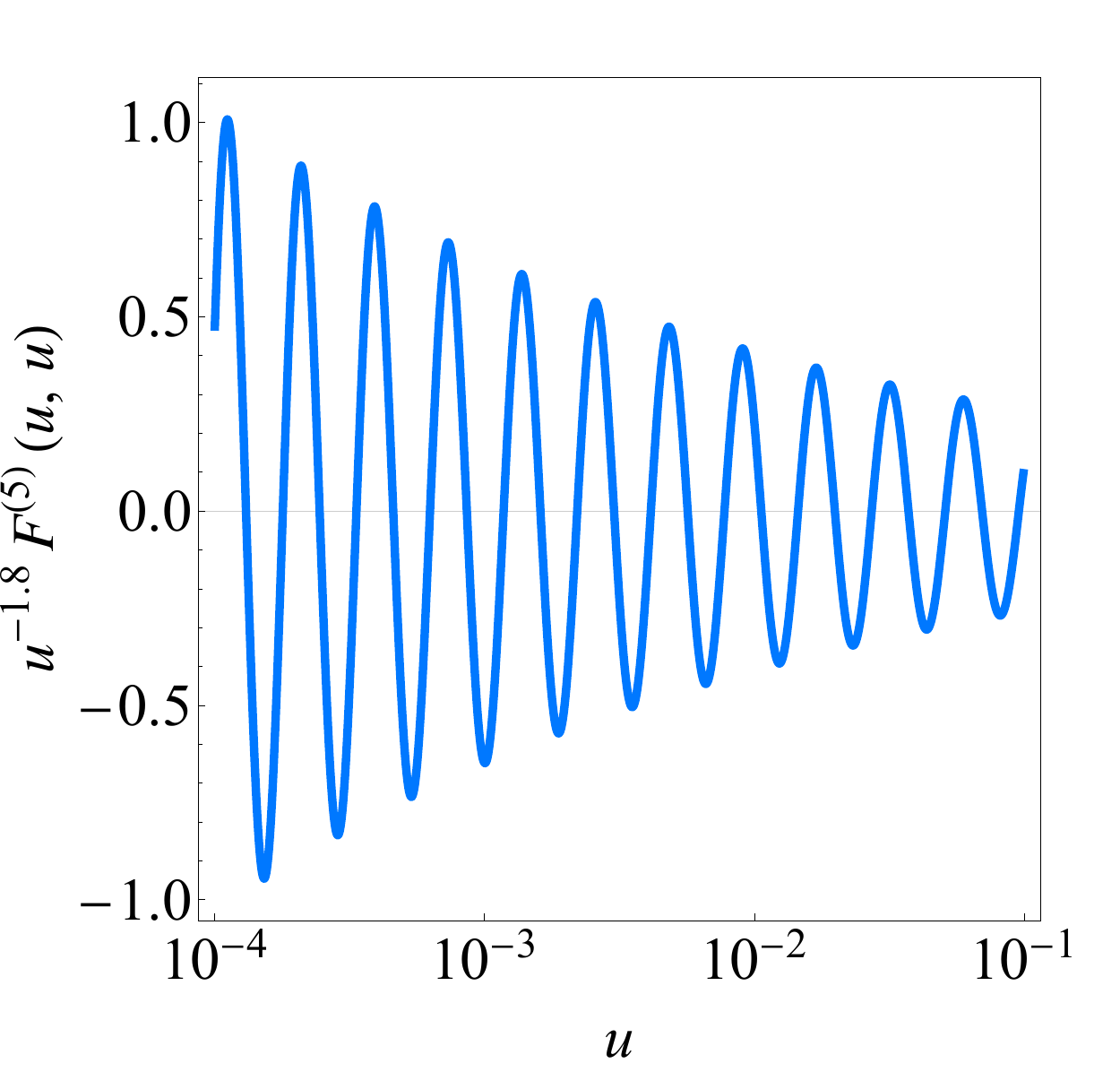}
        \caption{$\mu=5$}
    \end{subfigure}
    \caption{Behaviors in the collapsed limit when we fix $\Delta$ to be 1.8. In order to visually enhance the magnitudes for comparison, we rescaled $F^{(\mu)}(u,u)$ by $u^{-\Delta}$, and we set the absolute value of maxima/minima to be 1.}
    \label{fig:fixed_Delta_vary_mu}
\end{figure}

The Figure \ref{fig:fixed_Delta_vary_mu} shows the effect of $\mu$ when we fix $\Delta$. 
The frequency increases when the effective mass $\mu$ becomes larger. 
We should notice that the magnitude of the curves decreases very fast with the increase of $\mu$, from the Gamma function factors. 
In Figure \ref{fig:fixed_Delta_vary_mu}, we rescaled the functions by $u^{-\Delta}$ and rescaled the absolute value of maxima/minima to 1 for comparison. 

\begin{figure}[h]
    \centering
    \begin{subfigure}{0.45\textwidth}
        \includegraphics[width=\textwidth]{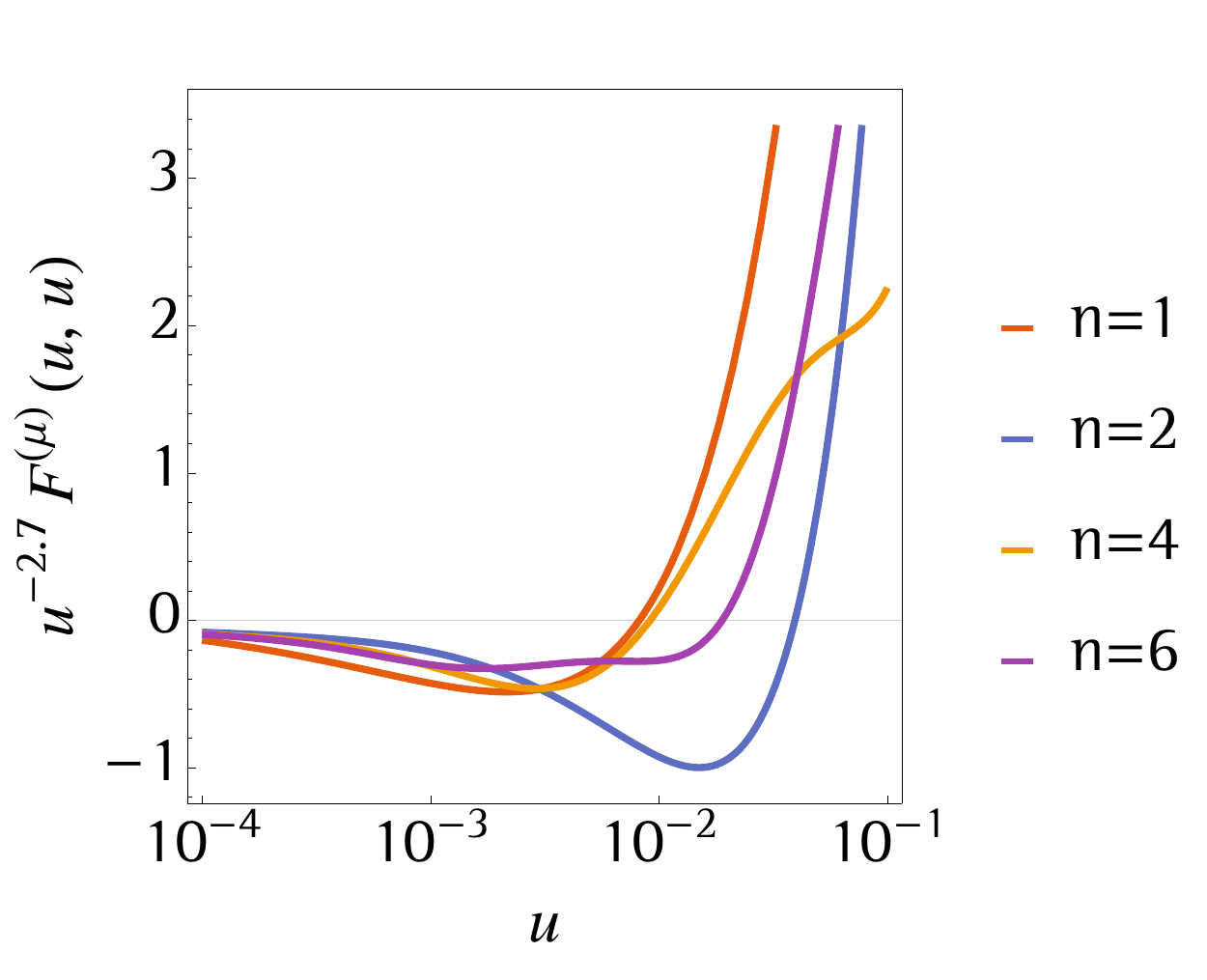}
        \caption{Localized interaction at $y_0=0$}
        \label{fig:localized_interaction}
    \end{subfigure}
    \hfill
    \begin{subfigure}{0.45\textwidth}
        \includegraphics[width=\textwidth]{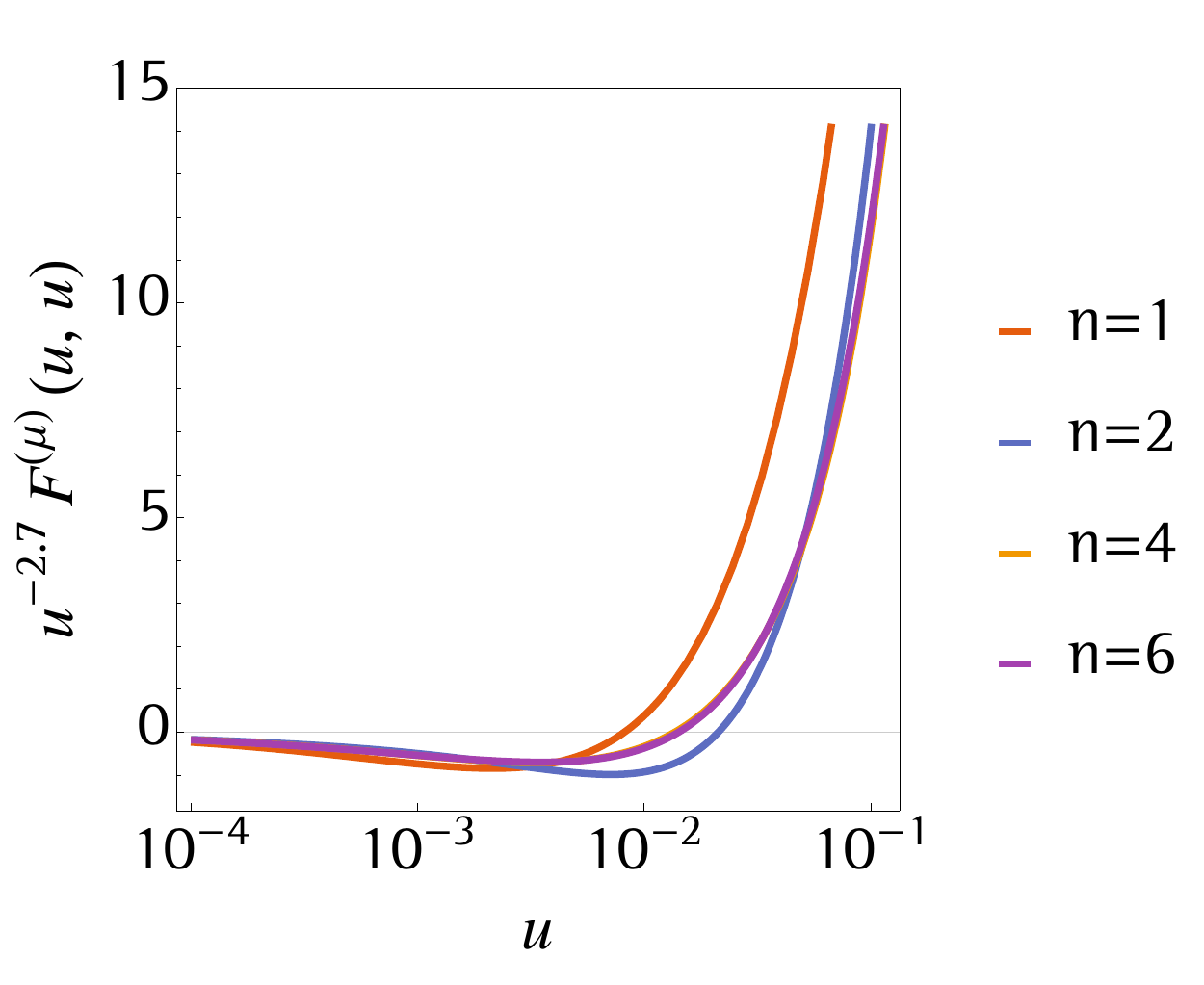}
        \caption{Spread interaction with $\kappa=4$ at $y_0=0$}
        \label{fig:spread_interaction}
    \end{subfigure}
    \caption{Interfering behaviors in the collapsed limit, when we fix $\Delta$ to $2.7$, and effective mass to $\mu=n/2\pi$. 
    The $n$'s in the legend denote the highest level of excitation in the KK tower we are considering. 
    The zero mode is not included. 
    In order to visually enhance the magnitudes for comparison, we rescaled $F^{(\mu)}(u,u)$ by $u^{-\Delta}$, and we rescaled the minima to $-1$.}
    \label{fig:interference}
\end{figure}

If we localize the interaction on $S^1$, we can construct a universal coupling between modes, as shown in \eqref{res:5d_brane_interaction}, \eqref{res:5d_fat_brane_interaction} and \eqref{def:cc_coupling}, and the modes will interfere. 
Assuming that inflation can effectively excite the tower of gapped unparticles up to level-$n$, Figure \ref{fig:interference} shows that in a certain window, how the modes interfere with localized and spread interaction with $y_0=0$. 
Away from the window, the interference will not be significant, since the function is suppressed by $\sim e^{-\pi\mu}$. 

Figure \ref{fig:spread_interaction} shows that a spread interaction will increase the degeneracy between different excitations. 
We can see that the curves for $n=4$ and $n=2$ are almost covering each other. 

\begin{figure}[h]
    \centering
    \includegraphics[width=0.5\linewidth]{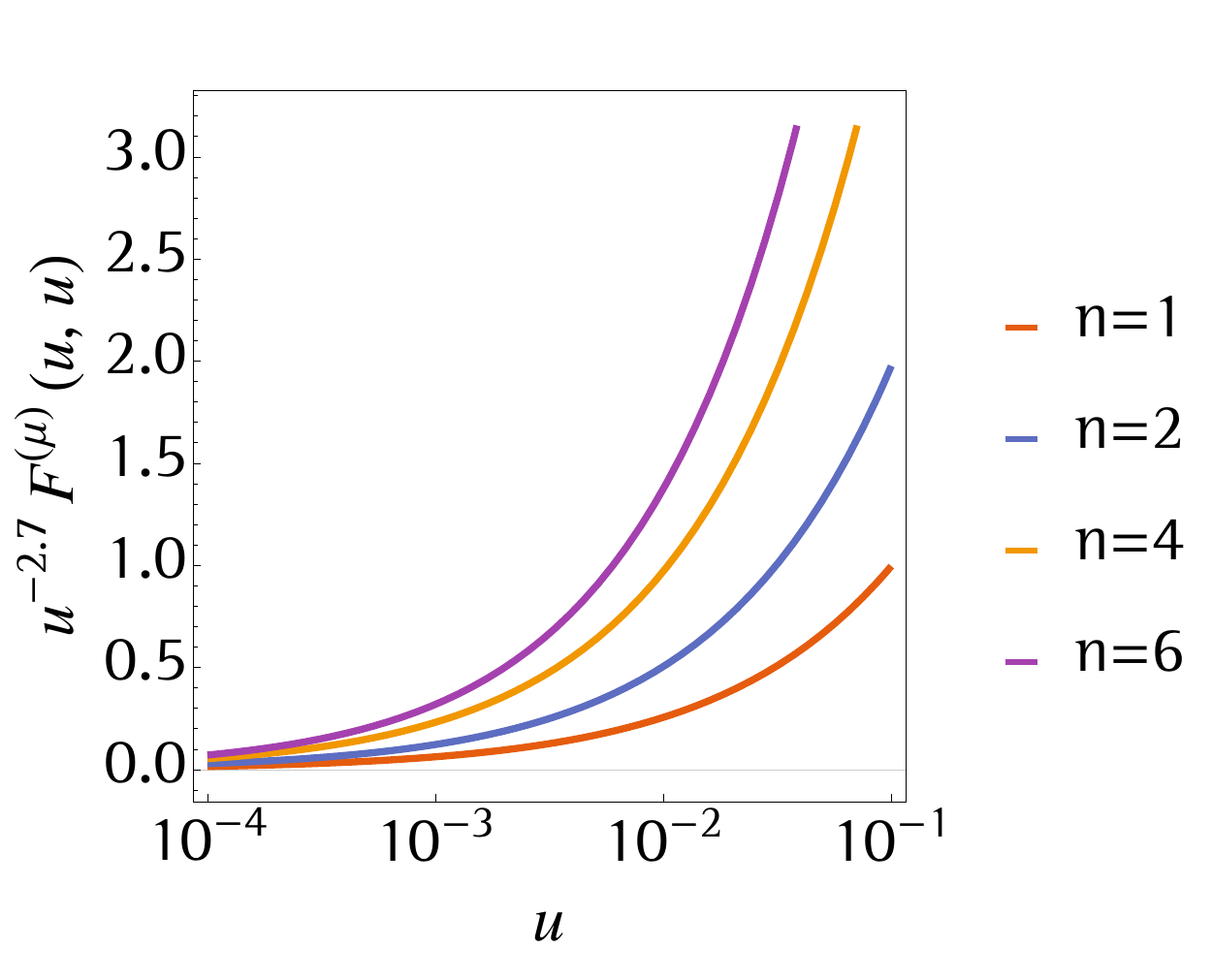}
    \caption{Behaviors in the collapsed limit with $\Delta=2.7$ and effective mass $\mu=n/100$. 
    The $n$'s in the legend denote the highest level of excitation in the KK tower we are considering. 
    The zero mode is not included. 
    In order to visually enhance the magnitudes for comparison, we rescaled $F^{(\mu)}(u,u)$ by $u^{-\Delta}$.}
    \label{fig:small_gap}
\end{figure}

If we consider a very small effective mass $\mu$, the spectrum will become dense and we expect the behaviors to be more similar to gapless unparticles. 
Figure \ref{fig:small_gap} shows that, if we take $HR=100$, even considering the interferences, the trispectra will lose the feature of oscillation in the collapsed limit. 

\begin{figure}[h]
    \centering
    \includegraphics[width=0.56\linewidth]{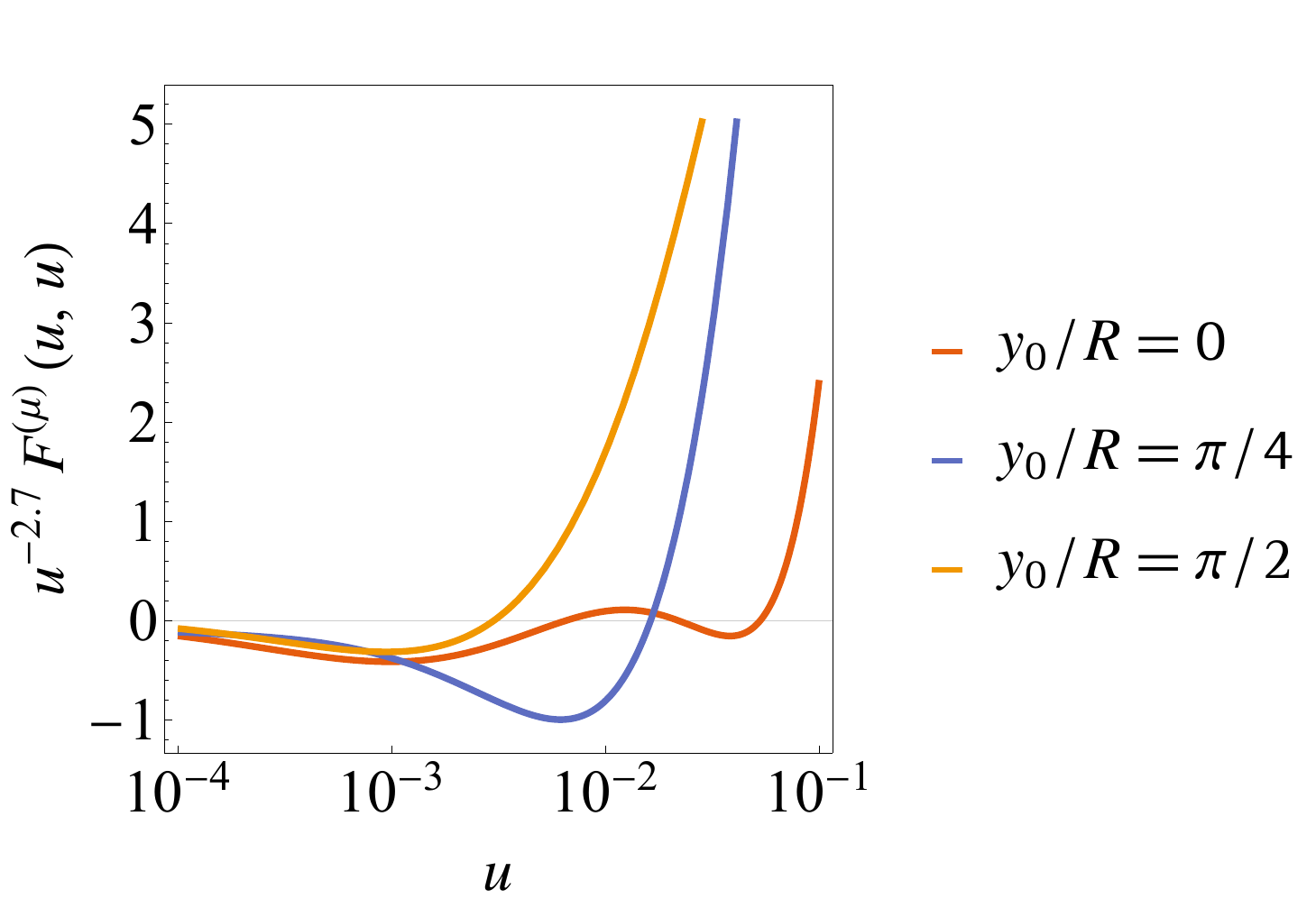}
    \caption{Behaviors in the collapsed limit with $\Delta=2.7$ and effective mass $\mu=n/2\pi$. 
    We fix the highest level of excitation in the KK tower to $n=3$. 
    The zero mode is not included. 
    In order to visually enhance the magnitudes for comparison, we rescaled $F^{(\mu)}(u,u)$ by $u^{-\Delta}$, and we rescaled the minima to $-1$.}
    \label{fig:brane_position}
\end{figure}

Figure \ref{fig:brane_position} shows three cases when we change $y_0$ of the localized interaction on $S^1$. 
The special feature of different $y_0$ is that, since $y_0$ introduces a phase factor $e^{iny_0/R}$ into the interactions of modes, it will project out certain parts for each mode. 
As a result, varying $y_0$ leads to selective enhancement or suppression of specific oscillatory components in the trispectrum. 
This provides a simple handle to continuously deform the detailed oscillatory patterns without altering the underlying mass spectrum. 

Here we only present a tiny bit of the parameter space to emphasize some of the features of gapped unparticle trispectra. 
The whole parameter space is controlled by the five-dimensional scaling dimension $\Delta$, the effective mass $\mu=n/(HR)$, the highest level of excitation $n$ and the initial relative phase $y_0/R$ of localized interactions. 
We only focused on the collapsed limit, it would be very interesting to also have analytic templates for the full shape. 

\begin{figure}[h]
    \centering
    \includegraphics[width=0.5\linewidth]{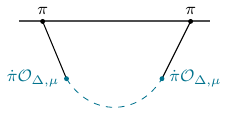}
    \caption{Tree-level diagram contribution to the power spectrum $\langle\pi\pi\rangle$. The solid and dashed lines represent the scalar perturbation $\pi$ and a gapped unparticle $\mathcal{O}_{\Delta,\mu}$ respectively.}
    \label{fig:correc_power_spectrum}
\end{figure}

Here we can also briefly discuss the constraints on the coupling from correction to the power spectrum, and then estimate for the achievable signal strength for trispectra\footnote{We thank the anonymous reviewer for bringing this up.}. Using the language of the EFT of inflation, the leading order interaction between the scalar perturbation $\pi$ and the gapped unparticle $\mathcal{O}_{\Delta,\mu}$ is 
\begin{align}
    &\mathcal{L}_{\pi\mathcal{O}} = \frac{1}{2}\alpha^{2-\Delta} M_{\text{pl}}|{\dot{H}}|^{1/2} \left(-2\dot{\pi}\mathcal{O}_{\Delta,\mu} + (\partial_\mu\pi)^2\mathcal{O}_{\Delta,\mu}\right),
\end{align}
where $\alpha$ is the effective coupling constant. As we can see, the quadratic interaction is what we described above for trispectra, and the linear mixing could give rise to a correction $\delta P$ of the power spectrum, as illustrated in Figure \ref{fig:correc_power_spectrum}.

At leading order, the curvature perturbation $\zeta$ relates to $\pi$ as 
\begin{align}
    \zeta=-H\pi+O(\pi^2). 
\end{align}
In slow-roll inflation model, it is natural to take $\alpha^{2-\Delta}\equiv\dot{\phi}/\Lambda^\Delta$, where $\Lambda$ is the UV scale to form the gapped unparticles. 
As an \textit{extreme} estimation, we will take the constraint on the corrections to power spectrum $P\equiv\langle\zeta\zeta\rangle$ to be $O(1)$: 
\begin{align}
    \frac{\delta P}{P} \propto (\alpha^{2-\Delta})^2 \lesssim 1 \Rightarrow \Lambda^\Delta\gtrsim\dot{\phi},
\end{align}
assuming the contribution from the time integrals of gapped unparticle to also be $O(1)$. 
The observational analysis \cite{Philcox:2025wts} gives the upper bound $\tau_{\rm NL}\lesssim1500$. 
If the trispectrum non-Gaussianity $\tau_{\rm NL}$ for a rhombus shape $k_1=k_2=k_3=k_4$ saturates the observation bound, combined with the constraint from the power spectrum, it will give constraints for $\Lambda$ to be $\Lambda\gtrsim10^{17}$ GeV and $\Delta>1.8$, if we take the contribution from the shape function to be $O(1)$, as typical values for $\mu\sim1/\pi$ and not too large $\Delta$. 
For larger $\Delta$, the contribution of the shape function decays fast. 
This is a reasonable UV scale for a potential dark sector, as it lies between Hubble scale and Planck mass.
\newpage
\section{Conclusions and Future Directions}
\label{sec:Outlook}

In this paper, we considered a gapped sector with large anomalous dimensions that couples weakly to the curvature perturbation. 
It is a modest construction from the dimensional reduction of an extra-dimensional circle, in the form of unparticles. 
An interesting extension of our work would be to consider strongly coupled theories with holographic duals. We make some general comments on the setup in Appendix \ref{app:holography}.
We have computed the late-time four-point correlator of conformally coupled scalars in de Sitter, in the collapsed limit, where the distinct signatures of these gapped unparticles are encoded. 
Unlike the gapless unparticle case \cite{Pimentel:2025rds}, the trispectra have novel oscillatory behaviors, depending on the five-dimensional scaling dimension $\Delta$ and the effective mass $\mu$. 

From the plots discussed in Section \ref{sec:comments_pheno}, for a single gapped unparticle mode, there is a clear strategy to distinguish gapped unparticles from heavy massive scalars in the \textit{Principal Series}. 
The frequency of the oscillation in the collapsed limit is controlled by the effective mass $\mu$, and the envelope of the oscillation is controlled by the five-dimensional scaling dimension $\Delta$. 
In this way, we can measure the parameters $\Delta$ and $\mu$ independently. 
Once we consider localized/spread interactions, universal couplings can be derived, and the unparticle modes will interfere. 
In a small window in the parameter space, significant interference up to different levels of the KK tower can be identified. 
The degeneracy between signatures of interference up to different excitations will increase, if the interactions spread along the brane. 
In this work, we only considered one peak of the localization. 
We also found that the initial relative phase of the localized interaction will change the oscillatory behaviors. 

We have only scratched some simple aspects of the gapped scenario, within a rich theoretical landscape of strongly coupled sectors in inflation, trying to open new avenues for future exploration. 
We provide some below: 
\begin{itemize}
    \item In this work, we study the gapped scenario. 
    Similar to the gapless scenario \cite{Pimentel:2025rds}, we still regard the strongly coupled sector as an independent sector, which is weakly coupled to inflatons. 
    In order to have a more complete UV completion, it is worth exploring to have a microscopic theory, which is able to give inflationary dynamics, visible sector and hidden sector at the same time. 
    \item We focus on the collapsed limit of trispectrum in this work. 
    It would be very interesting to have analytic templates valid at any kinematics for the trispectra. 
    \item In order to enhance non-Gaussianities, it is helpful to incorporate the effects of a small speed of sound, as discussed in \cite{Pimentel:2022fsc}. 
    Also in scenarios with a chemical potential \cite{Flauger:2016idt,Bodas:2020yho,Bodas:2025wuk}, non-Gaussianities can be enhanced because one can probe energy scales up to $\sim60H$ in slow-roll inflation. 
    How would the resulting shapes change in these cases? 
    \item The gapless unparticle has conformal symmetries, which give the controlling differential equations for the tree-level cosmological correlators.  
    Can we systematically develop a theory of differential equations for tree-level gapped unparticle diagrams?  
    \item We discussed the five-dimensional conformal field theory without specifying the underlying UV picture. 
    Can we develop practical models for gapped unparticles in the strongly coupled sector? 
    \item Finally, it would be interesting to constrain the parameters on gapped unparticles using observational data from the Cosmic Microwave Background (CMB) or Large Scale Structure (LSS) \cite{Philcox:2024jpd,Philcox:2025bvj,Philcox:2025lrr,El-Haj:2025zbe,Philcox:2025bbo,Sohn:2024xzd,Suman:2025tpv,Philcox:2026bfa}. 
\end{itemize}

Studying different possible mechanisms behind the primordial curvature perturbations might open new observational windows to probe them.  
The study of the mathematical structures for cosmological correlators will also improve our understanding of quantum field theories in cosmological spacetime. 
Much remains to be done in building an alternative viewpoint to slow-roll, ``Higgs-like" models of inflation, with distinct observational signatures.

\section*{Acknowledgements}
Thanks to Nima Arkani-Hamed, Craig Clark, Hao Geng, Ling-Yan Hung, Austin Joyce, Juan Maldacena, Raman Sundrum, Sonali Verma, Tom Westerdijk, and the Reviewer for useful discussions. A special thanks to Majid Ekhterachian for discussions and detailed comments on a draft.
CY thanks the Institute for Advanced Study, Tsinghua---in particular, Yixu Wang and Zhenbin Yang for hospitality. GLP thanks the Yukawa Institute
for Theoretical Physics at Kyoto University for hospitality during its “Progress of Theoretical
Bootstrap” workshop. He also thanks the Centro de Ciencias de Benasque Pedro Pascual for hospitality.

The work of YJ is supported by the U.S Department of Energy ASCR EXPRESS grant, Novel Quantum Algorithms from Fast Classical Transforms, and Northeastern University. 
GLP and CY are supported by Scuola Normale and by INFN (IS GSS-Pi). 
The research of GLP and CY is moreover supported by the ERC (NOTIMEFORCOSMO, 101126304). 
Views and opinions expressed are, however, those of the author(s) only and do not necessarily reflect those of the European Union or the European Research Council Executive Agency. 
Neither the European Union nor the granting authority can be held responsible for them. 
GLP is further supported by the Italian Ministry of Universities and Research (MUR) under contract 20223ANFHR (PRIN2022).

\appendix
\section{Calculation of von Mises Distribution}
\label{app:von_mises}

The von Mises distribution is the normal distribution on $S^1$, defined as 
\begin{align}
    f(x;x_0, \kappa) = \frac{e^{\kappa\cos(x-x_0)}}{2\pi I_0(\kappa)}, 
\end{align}
where $x_0$ is the measure of location, $\kappa$ is the measure of concentration and $2\pi I_0(\kappa)$ is the normalization factor. 
$I_{\nu}(z)$ is the modified Bessel function of the first kind. 

Now we compute the target integral in the following: 
\begin{align}
    &\int_{-\pi R}^{\pi R} \dint y\ \frac{e^{\kappa\cos(y/R-y_0/R)}}{2\pi I_0(\kappa)} e^{imy/R} \\
    =& \frac{R e^{imy_0/R}}{2\pi I_0(\kappa)} \int_{-\pi}^\pi \dint z\ e^{\kappa\cos(z-z_0)} e^{im(z-z_0)} \label{eqn:von_Mises_integral_before_shift} \\
    =& \frac{R e^{imy_0/R}}{2\pi I_0(\kappa)} \int_{-\pi}^\pi \dint z\ e^{\kappa\cos z} e^{imz} \label{eqn:von_Mises_integral_after_shift} \\
    =& \frac{R e^{imy_0/R}}{2\pi I_0(\kappa)} \times2\times \int_{0}^\pi \dint z\ e^{\kappa\cos z} \cos(mz) \\
    =& \frac{R e^{imy_0/R}}{I_0(\kappa)} I_m(\kappa). 
\end{align}
From \eqref{eqn:von_Mises_integral_before_shift} to \eqref{eqn:von_Mises_integral_after_shift}, we used the shift symmetry on $S^1$, and in the last step we used the integral representation of $I_{\nu}(z)$, which is 
\begin{align}
    I_\nu(z) = \frac{1}{\pi} \int_0^\pi \dint \theta\ e^{z\cos\theta} \cos(\nu\theta) - \frac{\sin(\nu\pi)}{\pi} \int_0^\infty \dint t\ e^{-z\cosh t-\nu t}, \quad \abs{\text{ph } z}<\pi/2. 
\end{align}
When the index $\nu\equiv m\in\mathds{Z}$, the second term on the right-hand-side vanishes and the representation simplifies to 
\begin{align}
    I_m(z) = \frac{1}{\pi} \int_0^\pi \dint \theta\ e^{z\cos\theta} \cos(m\theta). 
\end{align}

\section{Conformally Coupled Scalar in Five Dimensions}
\label{app:free_theory}

\begin{figure}[h]
    \centering
    \includegraphics[width=7cm]{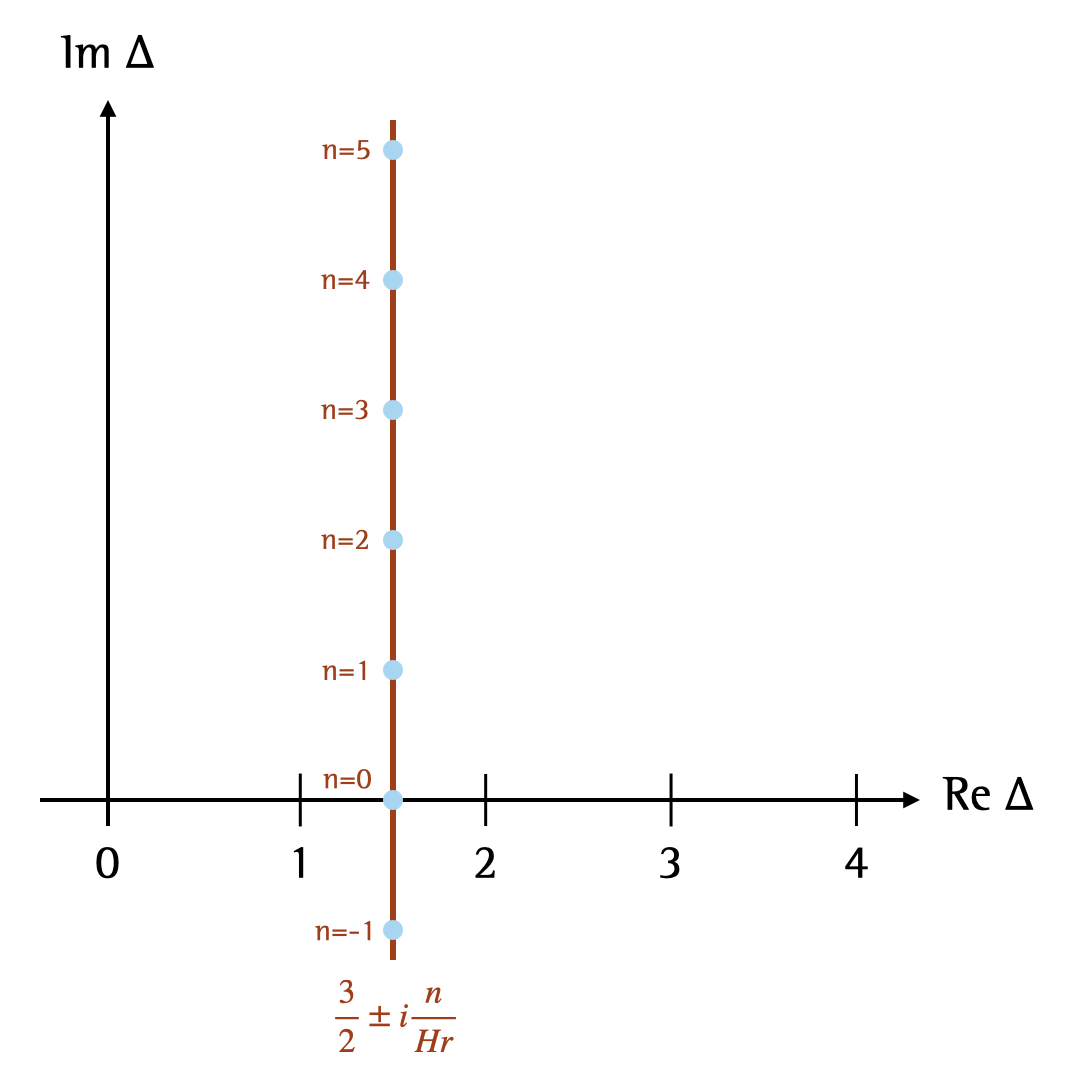}
    \caption{Boundary scaling dimension of four-dimensional effective modes}
    \label{fig:4d_modes}
\end{figure}

The simplest example for our setup is to consider a conformally-coupled scalar in five-dimensional geometry $\mathrm{dS}_4 \times S^1_R$. 
The conformally coupled scalar is defined as 
\begin{equation}
    S = -\frac{1}{2}\int \dint^D x \sqrt{-g} \Big(\partial_\mu\varphi\partial^\mu\varphi + \xi \mathrm{R} \varphi^2\Big), 
\end{equation}
where $\xi\equiv\frac{D-2}{4(D-1)}$, and $\mathrm{R}$ is the Ricci scalar. 
If we choose the metric to be \eqref{def:5D_metric}, the Ricci scalar is $\mathrm{R} = 12 H^2$. 
The equation of motion for conformally coupled scalars in the momentum space of \eqref{def:5D_metric} will be 
\begin{align}
    \Big(\eta^2\partial_\eta^2 - 2\eta\partial_\eta + k^2\eta^2 + \frac{9}{4} - \frac{1}{H^2} \partial_y^2\Big) \varphi = 0. 
    \label{eqn:eom_cc_scalar}
\end{align}
Expanding $\varphi$ in the $y$-direction, similar to \eqref{def:phi_5d_expansion}, will imply the equation of motion for the $n$-th Fourier mode in de Sitter: 
\begin{align}
    \Big(\eta^2\partial_\eta^2 - 2\eta\partial_\eta + k^2\eta^2 + \frac{9}{4} + \frac{n^2}{H^2R^2}\Big) \varphi_{n} = 0. 
    \label{eqn:4d_eff_eom}
\end{align}
Therefore, the effective four-dimensional scalars pick up an additional massive contribution from the windings on the circle. 
The three-dimensional boundary scaling dimension $\Delta$ \footnote{Readers should not get confused with the scaling dimension of unparticles in the main text.} for the four-dimensional effective theory is defined as 
\begin{align}
    \Delta(3-\Delta) \equiv m^2 = \frac{9}{4} + \frac{n^2}{H^2R^2}\ \Longrightarrow\ \Delta_{\pm} = \frac{3}{2} \pm i\frac{n}{HR}. 
\end{align}
This is exactly the \textit{Principal Series} of the unitary irreducible scalar representation in four-dimensional de Sitter space, as plotted in Figure \ref{fig:4d_modes}. 

The Principal Series usually refers to the \textit{heavy} scalars, which means we automatically generate heavy fields by performing KK reduction. 
\section{Weight-Shifting and Inflationary Trispectra}
\label{app:inf_tri}

The \textit{simplified} mode functions of conformally coupled scalars and inflatons are 
\begin{align}
    \varphi_k = \eta e^{-ik\eta+i\vec{k}\cdot\vec{x}}, \quad \phi_k = (1+ik\eta)e^{-ik\eta+i\vec{k}\cdot\vec{x}}. 
\end{align}
In \eqref{res:5d_brane_interaction}, \eqref{res:5d_fat_brane_interaction} and \eqref{def:4d_effective_action}, we couple the inflaton to unparticles in a shift-symmetric way. 
Thanks to the weight-shifting operator discussed in \cite{Arkani-Hamed:2015bza,Arkani-Hamed:2018kmz,Baumann:2019oyu}, we can relate the inflationary four-point functions to the four-point function of conformally coupled scalars by 
\begin{align}
    \nabla_\mu \phi_{k_1} \nabla^\mu \phi_{k_2} = s^2 U_{12}(\varphi_{k_1}\varphi_{k_2}), 
\end{align}
where the weight-shifting operator $U_{12}$ is 
\begin{align}
    U_{12}(\cdot)\equiv\frac{1}{2}\left(1-\frac{k_1 k_2}{k_{12}}\partial_{k_{12}}\right) \left(\frac{1-u^2}{u^2}\partial_{u}(u\cdot)\right), 
\end{align}
and $s$ is the Mandelstam-like variable $s\equiv|\vec{k}_1+\vec{k}_2|$. 
With the help of weight-shifting, the four-point functions $\mathcal{F}^{(\mu)}(u,v)$ for $\phi$ relates to $F^{(\mu)}(u,v)$ as 
\begin{align}
    \mathcal{F}^{(\mu)}(u,v) = s^3U_{12}U_{34}F^{(\mu)}(u,v). 
\end{align}

This procedure gives us a clear insight about the analytic structure of inflationary correlators. 
However, in the language of the EFT of inflation, $\nabla_\mu \phi_{k_1} \nabla^\mu \phi_{k_2}$ corresponds to $(\partial_\mu\pi)^2$, and the weight-shifting procedure restricts us to certain couplings between the Goldstone $\pi$ and the (gapped) unparticles $\mathcal{O}_\Delta$. 
To perform a more specific vertex matching, the de Sitter isometries needs to be more strongly broken, by giving a subluminal speed of propagation to the scalar fluctuations \cite{Pimentel:2022fsc}. 

\begin{figure}[h]
    \centering
    \includegraphics[width=0.6\linewidth]{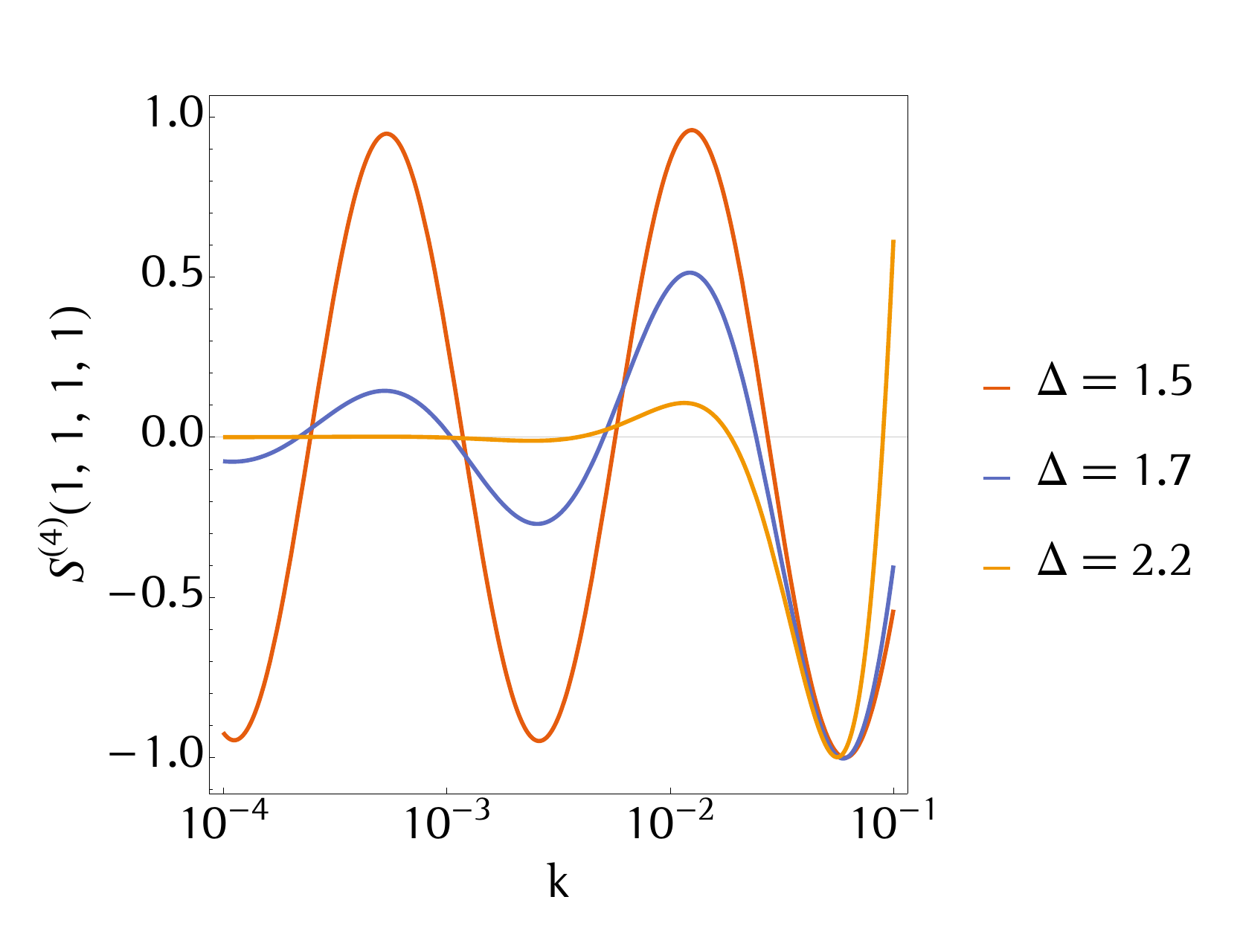}
    \caption{Trispectrum shape function in the collapsed limit when we fix $\mu\equiv1$, as examples. In order to visually enhance the magnitudes for comparison, we rescaled the minima to $-1$. We can see that when we fix the effective mass, the scaling dimension controls the envelope of the shapes.}
    \label{fig:inf_delta}
\end{figure}

\begin{figure}[h]
    \centering
    \begin{subfigure}{0.32\textwidth}
        \includegraphics[width=\textwidth]{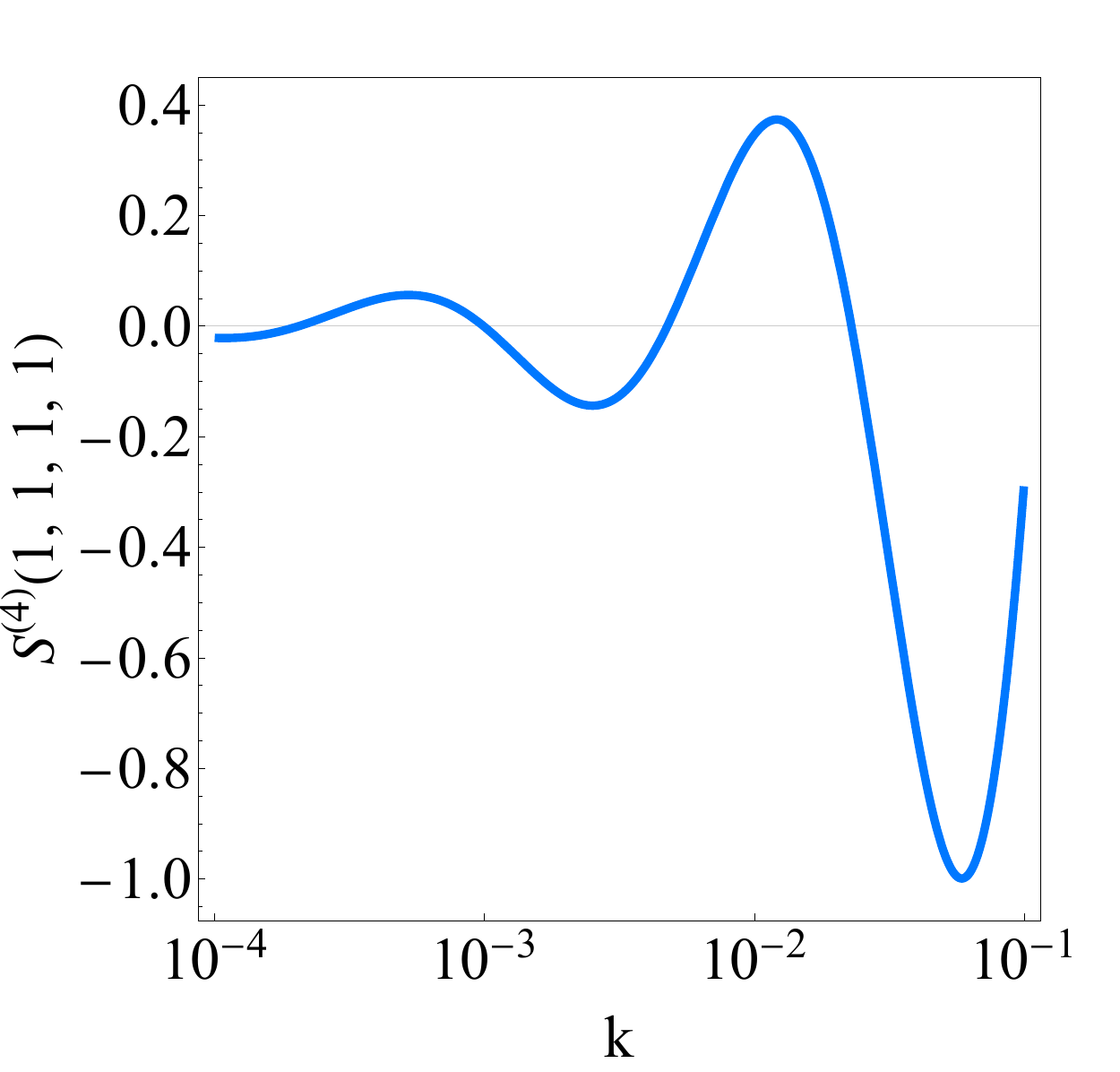}
        \caption{$\mu=1$}
    \end{subfigure}
    \hfill
    \begin{subfigure}{0.32\textwidth}
        \includegraphics[width=\textwidth]{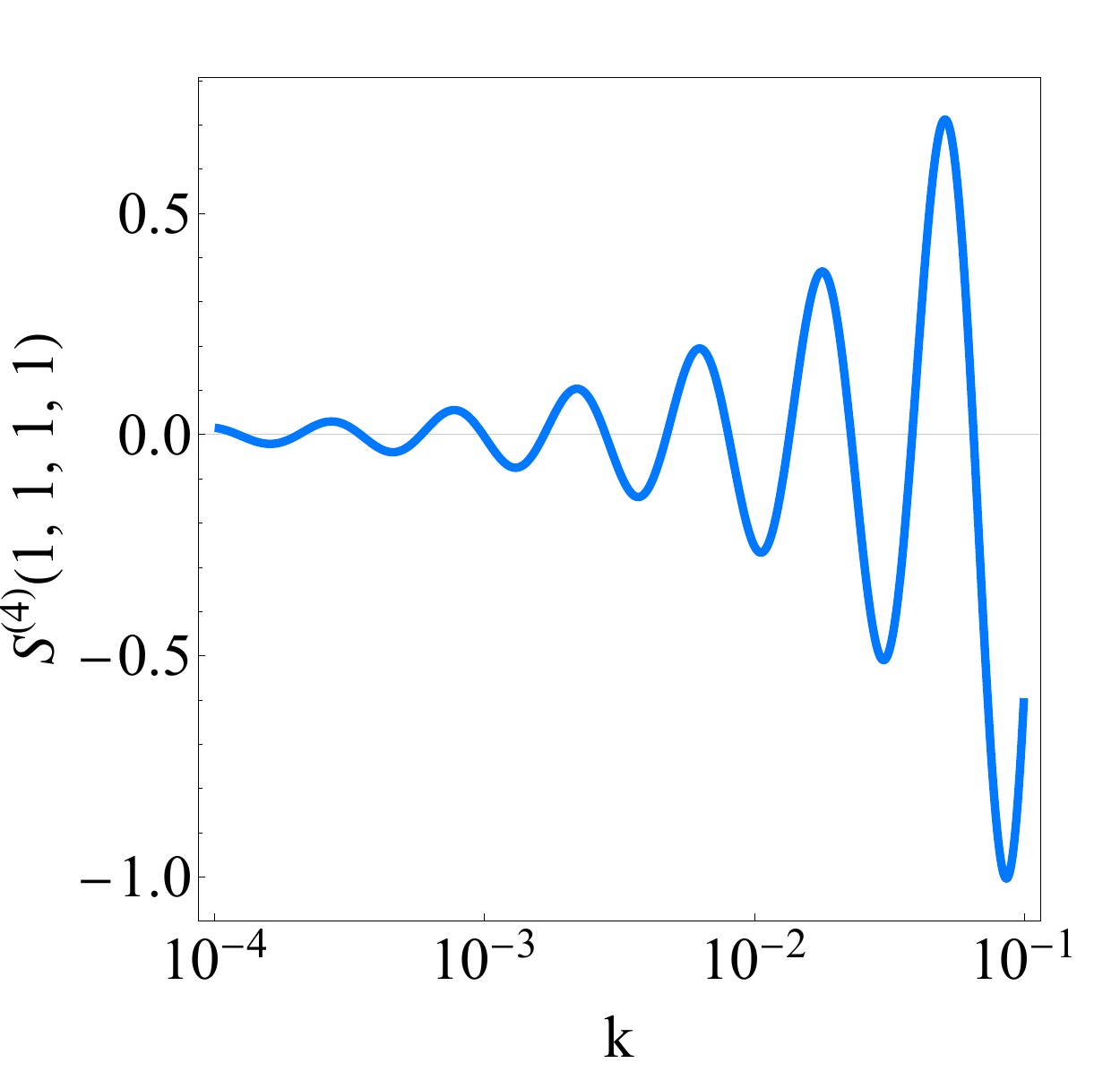}
        \caption{$\mu=3$}
    \end{subfigure}
    \hfill
    \begin{subfigure}{0.32\textwidth}
        \includegraphics[width=\textwidth]{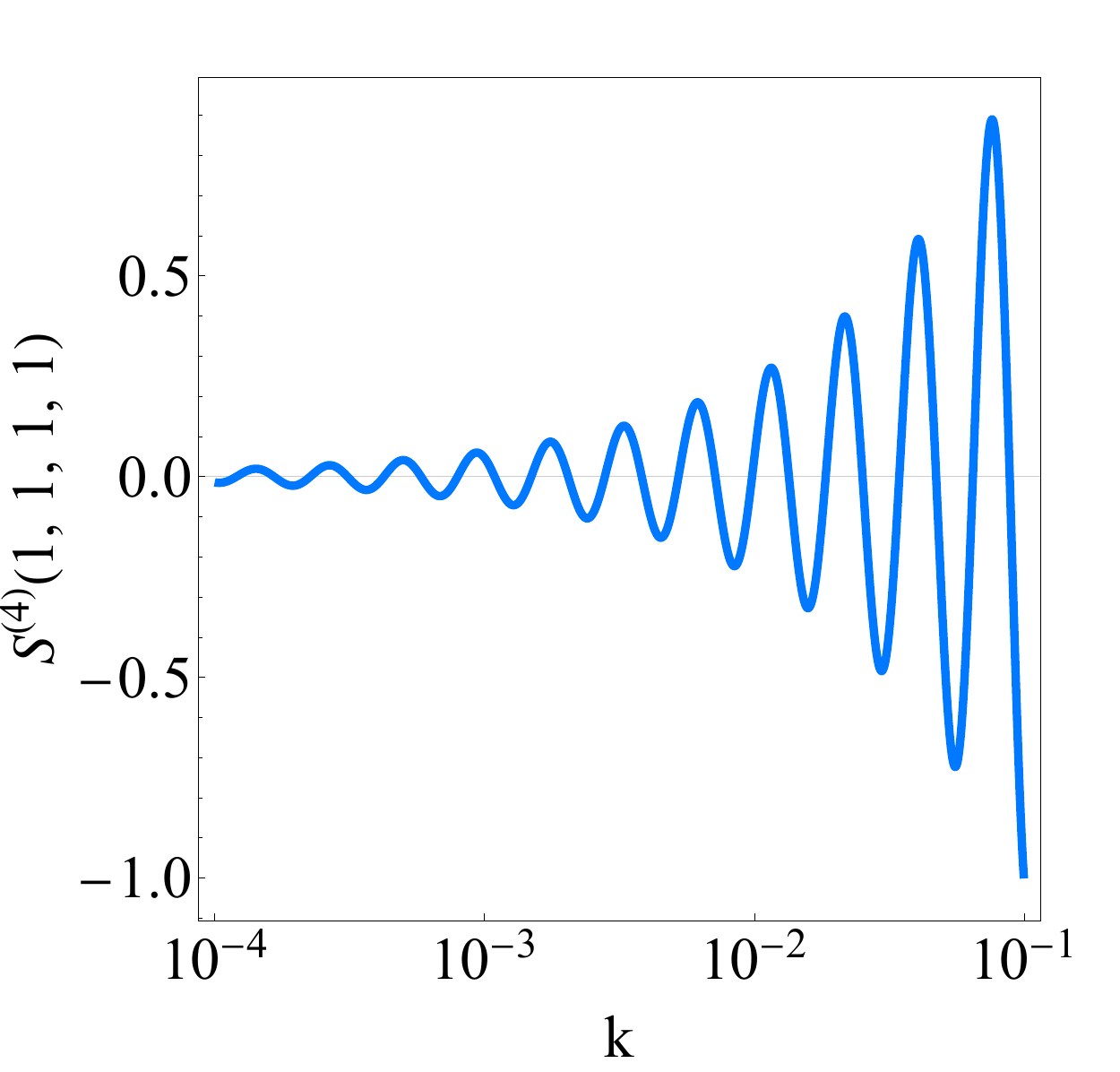}
        \caption{$\mu=5$}
    \end{subfigure}
    \caption{Trispectrum shape function in the collapsed limit when we fix $\Delta$ to be 1.8. In order to visually enhance the magnitudes for comparison, we rescaled the minima to $-1$. We can see that when we fix the scaling dimension, the effective mass controls the frequency of the oscillation.}
    \label{fig:inf_mu}
\end{figure}

We define the trispectrum to be 
\begin{align}
    \mathcal{T}_\phi (k_1,k_2,k_3,k_4) = s^3U_{12}U_{34}F^{(\mu)}(u,v) + \mathrm{permutations}, 
\end{align}
where the master function $F^{(\mu)}(u,v)$ is defined in \eqref{res:master_function}. 
We can define the dimensionless ``shape function" for trispectrum as 
\begin{align}
    S^{(4)}(k_1,k_2,k_3,k_4) \equiv \frac{k}{k_1k_2k_3k_4} \mathcal{T}_\phi (k_1,k_2,k_3,k_4)
\end{align}
To simplify the shape, we can just take $k_1=k_2=k_3=k_4=1$, choosing a rhombus shape, and the trispectrum shape function in the collapsed limit will be 
\begin{align}
    S^{(4)}(1,1,1,1) &= 3\times4^{-2\Delta-1-2i\mu} k^{2\Delta-2+2i\mu} \left(k^2(\Delta+3+i\mu)-4(\Delta+1+i\mu)\right)^2 (\Delta+i\mu)^2 \nonumber \\
    &\times \Big(1+\cos\big(\pi(\Delta+i\mu)\big)\Big) \Gamma(3/2-\Delta-i\mu) \Gamma(-i\mu) \Gamma(\Delta-1+i\mu)^2 \nonumber \\
    &+ 3\times4^{-2\Delta-1+2i\mu} k^{2\Delta-2-2i\mu} \left(k^2(\Delta+3-i\mu)-4(\Delta+1-i\mu)\right)^2 (\Delta-i\mu)^2 \nonumber \\
    &\times \Big(1+\cos\big(\pi(\Delta-i\mu)\big)\Big) \Gamma(3/2-\Delta+i\mu) \Gamma(i\mu) \Gamma(\Delta-1-i\mu)^2. 
\end{align}
We find that, the behaviors of the inflationary trispectrum shape function for $\Delta$ and $\mu$ parameters in the collapsed limit are not qualitatively different from what we discussed in the main text about conformally coupled scalars, as shown in Figure \ref{fig:inf_delta} and \ref{fig:inf_mu}.

\newpage
\section{Holographic Perspective}
\label{app:holography}

In previous sections, we introduced the gapped unparticles by performing dimensional reduction of a generalized free field/MFT on $\mathrm{dS}_4\times S^1_R$. 
If this five dimensional theory is strongly coupled, it is natural to consider a potential holographic dual.
Since now we have two scales, the Hubble scale $H$ and the radius $R$ of the circle, the competition between these scales will create two saddles in the six-dimensional geometry \cite{Witten:1998zw,Maldacena:2012xp,Marolf:2010tg}. 
When $HR>1$, the dominate one will be the \textit{topological black hole}, and the other one is the \textit{bubble of nothing}, when $HR<1$. 
The dominant geometry will be determined by the ratio between $H$ and $R$, which can be seen from the free energy. 
In the following, we will discuss these geometries in six dimensions. 
Our hope is that embedding this scenario in holography will give us access to more observables and richer phenomenology.

\subsection{Six-Dimensional Topological Black Hole---Gaps Around Hubble}
The geometry of an $\mathrm{AdS}_6$ with identification, as described in \cite{Banados:1997df, Banados:1998dc}, is called ``constant curvature black hole," or topological black hole. 
The metric defined outside the lightcone at the origin $r=0$ is 
\begin{equation}
    \dint s^2 = (1+H^2 r^2) \dint y^2 + H^2 r^2 \left(\frac{-\dint \eta^2 + \dint \vec{x}^2}{H^2\eta^2}\right) + \frac{\dint r^2}{1+H^2 r^2}. 
    \label{def:ads_ds_coord}
\end{equation}
This geometry has a five-dimensional $\text{dS}_4\times S^1_R$ slicing, defined in \eqref{def:5D_metric}. 
If we shift $\tilde{r}^2\equiv r^2+1/H^2$, the metric will become the classical form of constant curvature black holes \cite{Banados:1997df}: 
\begin{align}
    \dint s^2 = H^2 \tilde{r}^2 \dint y^2 + \left(H^2\tilde{r}^2-1\right) \left(\frac{-\dint \eta^2 + \dint \vec{x}^2}{H^2\eta^2}\right) + \frac{\dint \tilde{r}^2}{H^2\tilde{r}^2}. 
    \label{def:6d_bh_coord}
\end{align}
The horizon in this coordinate is obvious, $\tilde{r}_h=1/H$. 

\begin{figure}[h]
    \centering
    \includegraphics[width=0.5\linewidth]{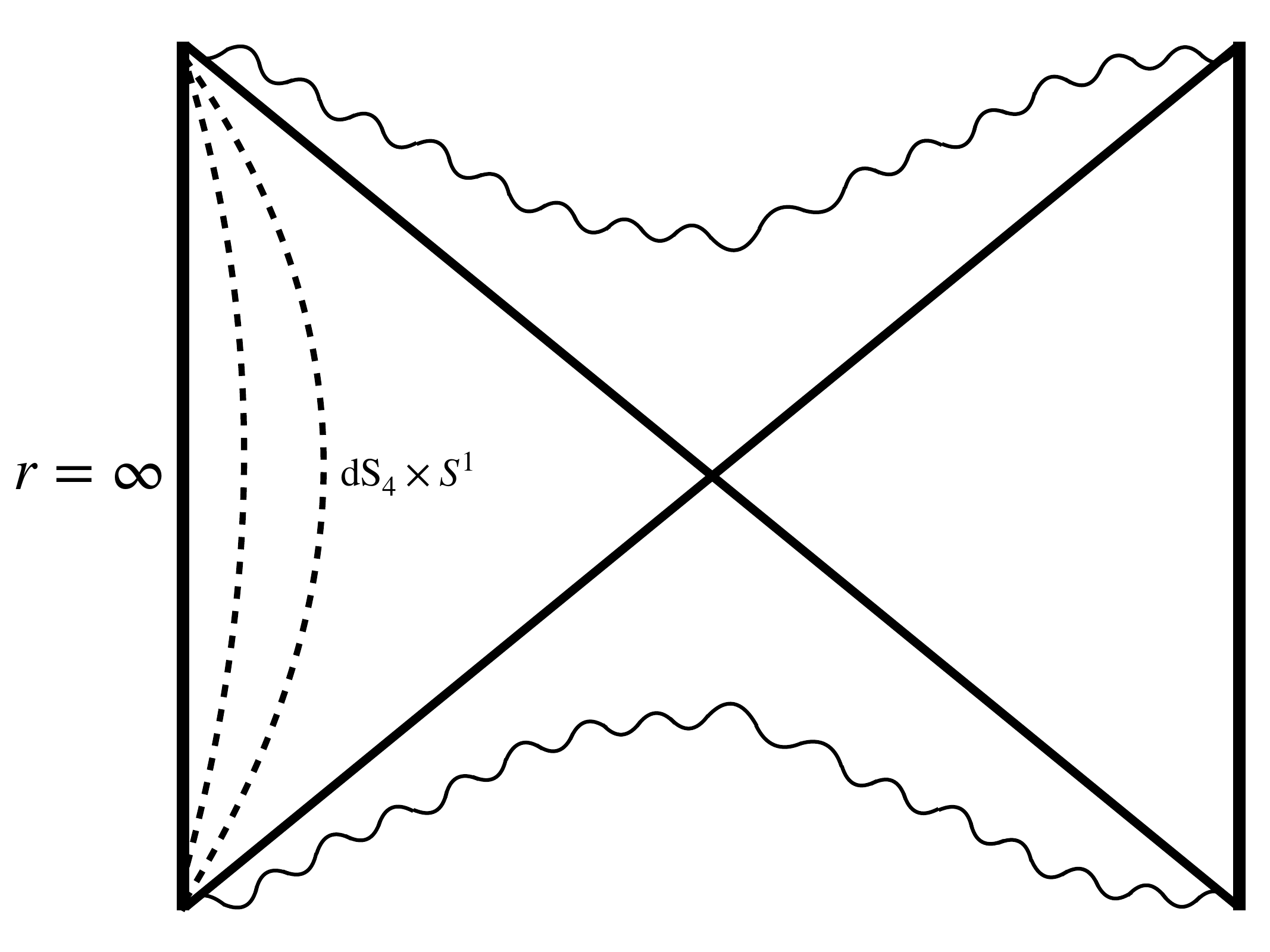}
    \caption{The Penrose diagram for six-dimensional topological black hole}
    \label{fig:topo_bh}
\end{figure}

Since the topological black holes can be defined via a periodic identification of vacuum AdS, the Green's function can be calculated by summing up the images. 
We can use the well-known AdS bulk-to-bulk propagator for free scalar with mass $M$ \cite{Burgess:1984ti,DHoker:2002nbb} by expressing chordal distance in terms of the coordinates in \eqref{def:ads_ds_coord}. 
To derive the dimensionless chordal distance $u\equiv(X-X')^2H^2/2$ in the embedding space, we need embedding relations, which are 
\begin{align}
    &\begin{aligned}
        & X^0 = \sqrt{r^2+1/H^2} \cosh(Hy), && X^2 = -\frac{r}{\eta} x_1, && X^5 = \frac{r}{2} \left(-H\eta-\frac{1}{H\eta}+H\frac{x_1^2+x_2^2+x_3^2}{\eta}\right), \\
        & && X^3 = -\frac{r}{\eta} x_2, && \\
        & X^1 = \sqrt{r^2+1/H^2} \sinh(Hy), && X^4 = -\frac{r}{\eta} x_3, && X^6 = \frac{r}{2} \left(H\eta-\frac{1}{H\eta}-H\frac{x_1^2+x_2^2+x_3^2}{\eta}\right), 
    \end{aligned} \label{eqn:bh_embedding_coord} \\
    & - (X^0)^2 + (X^1)^2 + (X^2)^2 + (X^3)^2 + (X^4)^2 + (X^5)^2 - (X^6)^2 = -1/H^2. 
\end{align}

The function $\xi$ related to the dimensional chordal distance $u$ is defined in \cite{DHoker:2002nbb}: 
\begin{align}
    \xi\equiv\frac{2}{1+u}. 
\end{align}
With embedding relations \eqref{eqn:bh_embedding_coord}, the chordal distance is 
\begin{align}
    u = \sqrt{1+H^2r_1^2} \sqrt{1+H^2r_2^2} \cosh(H(y_1-y_2))+H^2\frac{r_1 r_2 \left(-\eta_1^2-\eta_2^2+\left(\vec{x}_1-\vec{x}_2\right){}^2\right)}{2\eta_1\eta_2}-1. 
\end{align}

The six-dimensional AdS bulk-to-bulk scalar propagator in terms of $\xi$ is found to be 
\begin{align}
    G(\eta_1,\eta_2;\vec{x}_1,\vec{x}_2;y_1,y_2;r_1,r_2) = H^4\frac{C_{\Delta}}{2\nu} (\frac{\xi}{2})^{\Delta} {}_2F_1\left(\frac{\Delta}{2},\frac{\Delta+1}{2},\nu+1,\xi^2\right), 
\end{align}
where 
\begin{align}
    \Delta&\equiv\frac{5}{2}+\nu, \quad \nu\equiv\sqrt{\frac{M^2}{H^2}+\frac{25}{4}}, \quad C_{\Delta}\equiv\frac{\Gamma(\Delta)}{\pi^{5/2} \Gamma(\nu)}, \\
    \xi&=\frac{2\eta_1\eta_2}{H^2r_1 r_2 \left(-\eta_1^2-\eta_2^2+\left(\vec{x}_1-\vec{x}_2\right)^2\right)+2\eta_1\eta_2\sqrt{1+H^2r_1^2}\sqrt{1+H^2r_2^2}\cosh(H(y_1-y_2))}. 
\end{align}
Regarding the identification $(y_1-y_2) \to (y_1-y_2)+m\beta$, the $m$-th winding of Green's function is 
\begin{align}
    &G_m(\eta_1,\eta_2;\vec{x}_1,\vec{x}_2;y_1,y_2;r_1,r_2) = H^4\frac{C_{\Delta}}{2\nu} (\frac{\xi_m}{2})^{\Delta} {}_2F_1\left(\frac{\Delta}{2},\frac{\Delta+1}{2},\nu+1,\xi_m^2\right), \\
    &\xi_m=\frac{2\eta_1\eta_2}{H^2r_1 r_2 \left(-\eta_1^2-\eta_2^2+\left(\vec{x}_1-\vec{x}_2\right)^2\right)+2\eta_1\eta_2\sqrt{1+H^2r_1^2}\sqrt{1+H^2r_2^2}\cosh(H(y_1-y_2+m\beta))}. 
\end{align}
Taking the asymptotic limit $r_1,r_2\to \infty$, the corresponding boundary-to-boundary propagator will be 
\begin{align}
    &\langle\mathcal{O}_{\Delta} (\eta_1,\vec{x}_1,y_1) \mathcal{O}_{\Delta} (\eta_2,\vec{x}_2,y_2)\rangle_m = H^{2\Delta-4} \frac{2\nu}{C_\Delta}\lim_{r_1,r_2\to \infty} \left(H^2r_1 r_2\right)^{\Delta} G_m(\eta_1,\eta_2;\vec{x}_1,\vec{x}_2;y_1,y_2;r_1,r_2) \nonumber \\
    &= \left(\frac{H^2\eta_1\eta_2}{-\eta_1^2-\eta_2^2+\left(\vec{x}_1-\vec{x}_2\right)^2 + 2\eta _1\eta _2 \cosh\left(H(y_1-y_2+m\beta)\right)}\right)^\Delta. 
\end{align}
This is the same as the $m$-th mode in \eqref{res:5d_2pt_gapped_cft}. 

\subsection{Six-Dimensional Bubble of Nothing---Large Gap Limit}
\label{sec:cigar_geo}

When the AdS radius is larger than the circle radius, $HR<1$, the circle shrinks to zero before AdS reaches the singularity. 
The geometry will end at a finite distance. 
This is called \textit{Bubble of Nothing}, and is a typical holographic dual of gapped systems \cite{Balasubramanian:2002am,Balasubramanian:2005bg,Maldacena:2012xp}. 
The metric is given by 
\begin{align}
    \dint s^2 &= f\dint y^2 + H^2r^2\dint s^2_{\text{dS}_4} + \frac{\dint r^2}{f}, \quad f \equiv 1+H^2 r^2-\frac{r_0}{H^3 r^3} \\
    &= \Big(1+H^2 r^2-\frac{r_0}{H^3 r^3}\Big) \dint y^2 + H^2 r^2 \Big(\frac{-\dint \eta^2 + \dint \vec{x}^2}{H^2 \eta^2}\Big) + \frac{\dint r^2}{1+H^2 r^2-\frac{r_0}{H^3 r^3}}. 
    \label{def:6d_cigar_metric}
\end{align}
$r_0$ is a dimensionless parameter that can be written as 
\begin{align}
    r_0=H^3r_h^3(1+H^2r_h^2), 
    \label{def:r_0}
\end{align}
where $r_h$ is the largest root of $f(r)=0$, denoting the end of the geometry. 
The period $\beta$ of $y$-circle is now constrained to 
\begin{equation}
    \beta \equiv \frac{4\pi}{f'(r_h)} = \frac{4\pi r_h}{3+5H^2r_h^2}, 
    \label{def:periodicity}
\end{equation}
in the request of smoothness of geometry \cite{Witten:1998zw,Balasubramanian:2002am}. 

From the expression we can see that $\beta_{\text{max}} = 2\pi/(\sqrt{15}H)$ and this solution only exists for $\beta\leq\beta_{\text{max}}$. 
\begin{figure}[h]
    \centering
    \includegraphics[width=0.5\linewidth]{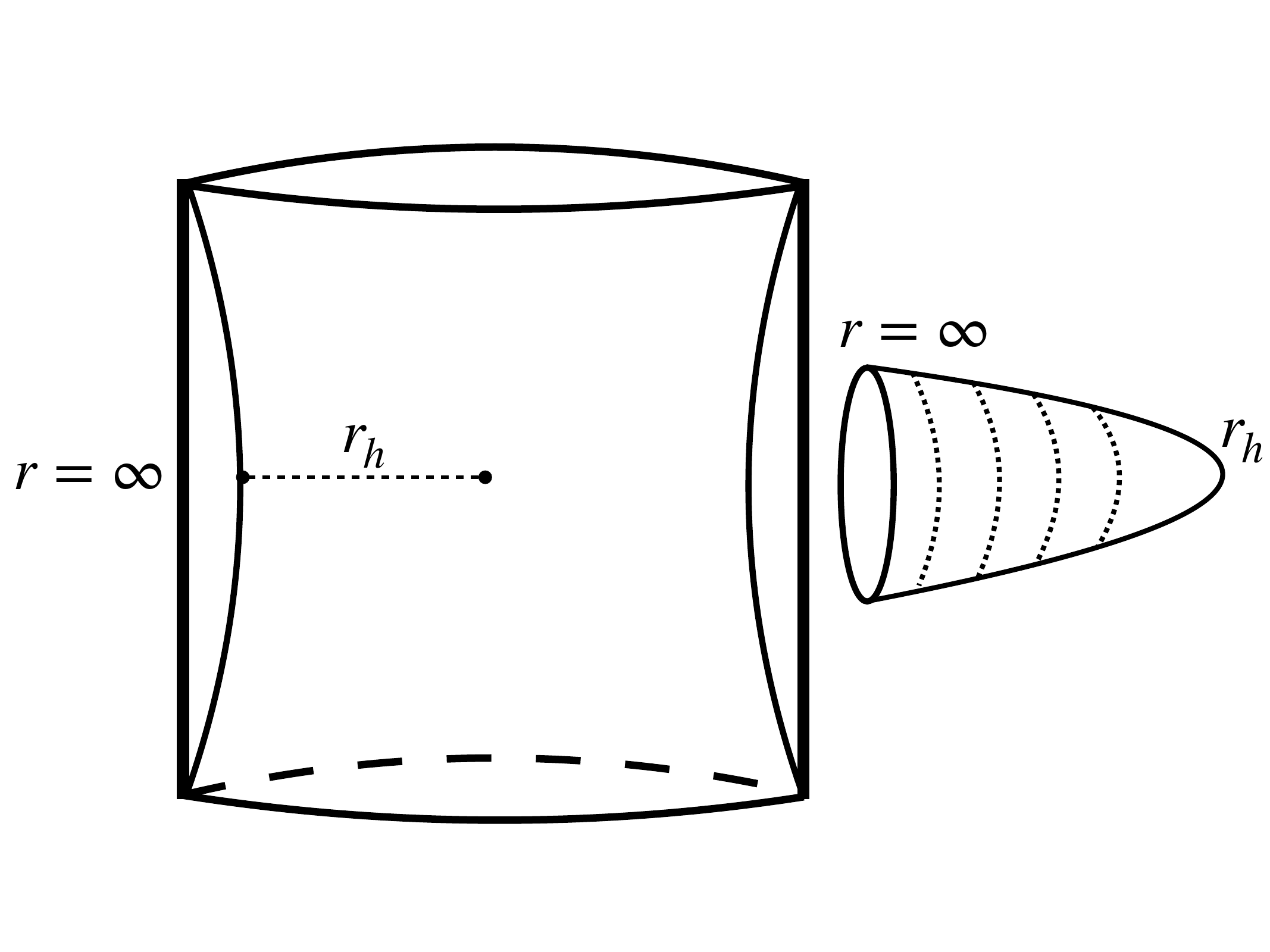}
    \caption{The six-dimensional cigar geometry for bubble of nothing}
    \label{fig:cigar_geo}
\end{figure}

For future simplicity, it is useful to denote the radial coordinate with $z\equiv1/r$. 
The equation of motion for free scalars $\Phi$ with mass $\mu$ in metric \eqref{def:6d_cigar_metric} is 
\begin{align}
    & \Box\Phi = \mu^2 \Phi \\
    \Rightarrow& \left(2\eta\partial_\eta - \eta^2\partial_\eta^2 + \eta^2\partial_{\vec{x}}^2\right)\Phi + \left(\frac{1}{z^2f(z)} \partial_y^2 + z^2f(z) \partial_z^2 + \left(z^2f'(z)-2zf(z)\right)\partial_z\right) \Phi - \frac{\mu^2}{z^2} \Phi = 0. 
\end{align}
It is good to see that the equation of motion separates into two parts, which means we can make an ansatz that the solution is factorized into 
\begin{equation}
    \Phi(\eta,\vec{x},y,z) \equiv I(\eta,\vec{x}) K(y,z). 
\end{equation}
Therefore, we will have a pair of equations, where the separation constant is chosen to be $M^2/H^2$, 
\begin{align}
    & \left(2\eta\partial_\eta - \eta^2\partial_\eta^2 + \eta^2\partial_{\vec{x}}^2\right) I(\eta,\vec{x}) \equiv \frac{M^2}{H^2} I(\eta,\vec{x}), \label{eqn:eom_cigar_I} \\
    & \frac{1}{z^2} \left(\frac{1}{f(z)} \partial_y^2 + z^4f(z) \partial_z^2 + \left(z^4f'(z)-2z^3f(z)\right)\partial_z - \mu^2\right) K(y,z) \equiv -\frac{M^2}{H^2} K(y,z). \label{eqn:eom_cigar_K}
\end{align}
\eqref{eqn:eom_cigar_I} is exactly the equation of motion for a free scalar with mass $M$ in $\text{dS}_4$. 
The mode function in momentum space for \eqref{eqn:eom_cigar_I} will be 
\begin{equation}
    I_k(\eta) = \frac{H\sqrt{\pi}}{2} e^{i\pi/4} e^{-\pi\sigma/2} (-\eta)^{3/2} H^{(1)}_{i\sigma}(-k\eta), \quad \text{where } \sigma\equiv\sqrt{\frac{M^2}{H^2}-\frac{9}{4}}, 
\end{equation}
in Bunch-Davies vacuum. 
We expand $K(y,z)$ in Fourier modes along $y$-direction to perform dimensional reduction 
\begin{align}
    K(y,z) \equiv \sum_{m=-\infty}^\infty K_m(z)\ e^{imy/\beta}. 
\end{align}
The equation for $K(y,z)$ now becomes 
\begin{align}
    &\left(\partial_z^2+\left(\frac{f'(z)}{f(z)}-\frac{2}{z}\right) \partial_z - \frac{1}{z^2f(z)} \left(\frac{\mu^2}{z^2}-\frac{M^2}{H^2}+\frac{1}{z^2f(z)}\frac{m^2}{\beta^2}\right)\right) K_m(z)=0. \label{eqn:K_m_mode}
\end{align}
We want to perform standard WKB approximation for \eqref{eqn:K_m_mode}, in the large $r_0$ limit. 
When $r_0$ is large, the value for the position of bubble $r_h$ will be large based on \eqref{def:r_0}, and $\beta$ will be small from \eqref{def:periodicity}. 
Considering the regime where we can neglect the backreaction of massive scalars on the background geometry, we require the parameters $\mu$, $M$ and $1/\beta$ to be comparably large, and take the ansatz $K_m(z)=\exp{(1/\delta\sum_n\delta^n\alpha_n(z))}$. 
$\delta$ is defined to be $\delta\ll1$, and we denote $\mu\equiv\bar{\mu}/\delta$, $M\equiv\bar{M}/\delta$ and $\beta\equiv\bar{\beta}\delta$, where $\bar{\mu}$, $\bar{M}$ and $\bar{\beta}$ are $O(1)$ parameters. 
The equation becomes 
\begin{align}
    &\left(\partial_z^2+\left(\frac{f'(z)}{f(z)}-\frac{2}{z}\right) \partial_z - \frac{1}{z^2f(z)} \left(\frac{\bar{\mu}^2}{\delta^2z^2}-\frac{\bar{M}^2}{\delta^2H^2}+\frac{1}{\delta^2z^2f(z)}\frac{m^2}{\bar{\beta}^2}\right)\right) e^{\frac{1}{\delta}\sum_n\delta^n\alpha_n(z)}=0. 
\end{align}
At the leading order, there are two solutions: 
\begin{align}
    \alpha_{0}'(z) &\rightarrow \frac{1}{z\sqrt{f(z)}}\sqrt{\frac{\bar{\mu}^2}{z^2} - \frac{\bar{M}^2}{H^2} + \frac{1}{z^2f(z)}\frac{m^2}{\bar{\beta}^2}}, & &(+)\text{-branch} \\
    \alpha_{0}'(z) &\rightarrow -\frac{1}{z\sqrt{f(z)}}\sqrt{\frac{\bar{\mu}^2}{z^2} - \frac{\bar{M}^2}{H^2} + \frac{1}{z^2f(z)}\frac{m^2}{\bar{\beta}^2}}, & &(-)\text{-branch}. 
\end{align}
The next order is identical for both branches, 
\begin{align}
    \alpha_1(z) = 2\log z - \frac{1}{4}\log\left(m^2H^2-\bar{M}^2z^2\bar{\beta}^2f(z) + \bar{\mu}^2H^2\bar{\beta}^2f(z)\right). 
\end{align}
Collecting the results above, the solution of \eqref{eqn:K_m_mode} up to second order in WKB approximation is 
\begin{align}
    K_m(z) &= \frac{z^2}{\left(m^2H^2-\bar{M}^2z^2\bar{\beta}^2f(z) + \bar{\mu}^2H^2\bar{\beta}^2f(z)\right)^{1/4}} \Bigg(C_1 \exp\left(\int_z^{z_h}\frac{\dint z}{z\sqrt{f(z)}}\sqrt{\frac{\mu^2}{z^2} - \frac{M^2}{H^2} + \frac{1}{z^2f(z)}\frac{m^2}{\beta^2}}\right) \nonumber \\
    &+ C_2 \exp\left(-\int_z^{z_h}\frac{\dint z}{z\sqrt{f(z)}}\sqrt{\frac{\mu^2}{z^2} - \frac{M^2}{H^2} + \frac{1}{z^2f(z)}\frac{m^2}{\beta^2}}\right)\Bigg), 
    \label{eqn:K_m_after_WKB}
\end{align}
where $z_h$ is the end of geometry in $z$-coordinate. 
To have \eqref{eqn:K_m_after_WKB} vanish when $z = z_h$, we require $C_2=-C_1$ and normalize $C_2$ to be 1. 

In the asymptotic limit $z\rightarrow0$, the solution \eqref{eqn:K_m_after_WKB} will become 
\begin{align}
    &\begin{aligned}
        K_m(z) =& c_0 \Bigg(\exp\left(-\int_z^{z_h}\frac{\dint z}{z\sqrt{f(z)}}\sqrt{\frac{\mu^2}{z^2} - \frac{M^2}{H^2} + \frac{1}{z^2f(z)}\frac{m^2}{\beta^2}}\right) \\
        -&\exp\left(\int_z^{z_h}\frac{\dint z}{z\sqrt{f(z)}}\sqrt{\frac{\mu^2}{z^2} - \frac{M^2}{H^2} + \frac{1}{z^2f(z)}\frac{m^2}{\beta^2}}\right)\Bigg),       
    \end{aligned} \\
    & c_0 \equiv \frac{z^{5/2}}{\sqrt{H\beta}\sqrt{H\mu}}. 
\end{align}
There seems to exist a divergence in the integral near $z\to0$. 
However, this can be removed by rewriting the expression as 
\begin{align}
    \begin{aligned}
        K_m(z) &= c_0 \exp\left(-\int_z^{z_h}\frac{\mu}{Hz}\right) \exp\left(-\int_z^{z_h}\dint z \left(\frac{1}{z\sqrt{f(z)}}\sqrt{\frac{\mu^2}{z^2} - \frac{M^2}{H^2} + \frac{1}{z^2f(z)}\frac{m^2}{\beta^2}} - \frac{\mu}{Hz}\right) \right) \\ 
        & - c_0 \exp\left(\int_z^{z_h}\frac{\mu}{Hz}\right)\exp\left(\int_z^{z_h}\dint z \left(\frac{1}{z\sqrt{f(z)}}\sqrt{\frac{\mu^2}{z^2} - \frac{M^2}{H^2} + \frac{1}{z^2f(z)}\frac{m^2}{\beta^2}} - \frac{\mu}{Hz}\right) \right). 
    \end{aligned}
\end{align}
The second exponential terms are finite now, since we have subtracted the superficially divergent pieces. 
The final result for $K_m$ after regularization is 
\begin{align}
    \begin{aligned}
        K_m(z) &= \frac{z_h^{-\mu/H} z^{5/2+\mu/H}}{\sqrt{H\beta}\sqrt{H\mu}} \exp\left(-\int_z^{z_h}\dint z \left(\frac{1}{z\sqrt{f(z)}}\sqrt{\frac{\mu^2}{z^2} - \frac{M^2}{H^2} + \frac{1}{z^2f(z)}\frac{m^2}{\beta^2}}-\frac{\mu}{Hz}  \right) \right)\\
        &- \frac{z_h^{\mu/H} z^{5/2-\mu/H}}{\sqrt{H\beta}\sqrt{H\mu}}  \exp\left(\int_z^{z_h}\dint z \left(\frac{1}{z\sqrt{f(z)}}\sqrt{\frac{\mu^2}{z^2} - \frac{M^2}{H^2} + \frac{1}{z^2f(z)}\frac{m^2}{\beta^2}}-\frac{\mu}{Hz}  \right) \right). 
    \end{aligned}
\end{align}
Based on the AdS/CFT dictionary \cite{Maldacena:1997re,Witten:1998qj,Klebanov:1999tb,Hartnoll:2016apf,Christodoulou:2017ceo}, the two-point function on the boundary contributed from the asymptotic behavior in $z\to0$ is 
\begin{align}
    &\langle\mathcal{O}_{\Delta}(\eta_1,k)\mathcal{O}_{\Delta}(\eta_2,k)\rangle_m \nonumber \\
    =& -H(2\Delta-5) \frac{\Phi_{(1)}(z)}{\Phi_{(0)}(z)}\Bigg|_{z\to0} \left(I_{k}(\eta_1)I_{k}^*(\eta_2)\Theta(\eta_1-\eta_2) + I_{k}^*(\eta_1)I_{k}(\eta_2)\Theta(\eta_2-\eta_1)\right) \\
    =&\ 2\mu\ z_h^{-2\mu/H} \exp\left(-2\int_z^{z_h}\dint z \left(\frac{1}{z\sqrt{f(z)}}\sqrt{\frac{\mu^2}{z^2} - \frac{M^2}{H^2} + \frac{1}{z^2f(z)}\frac{m^2}{\beta^2}}-\frac{\mu}{Hz}\right)\right) \nonumber \\
    \times& \left(I_{k}(\eta_1)I_{k}^*(\eta_2)\Theta(\eta_1-\eta_2) + I_{k}^*(\eta_1)I_{k}(\eta_2)\Theta(\eta_2-\eta_1)\right). 
    \label{res:large_gap_2pt}
\end{align}

We can investigate a bit on how much the two-point function is suppressed. 
For simplicity, from now we take $H\equiv1$. 
Taking the large $r_0$ limit, we write $z_h\sim\delta$. 
Thus in the leading order, $r_0\sim1/\delta^5$ from \eqref{def:r_0}, $\beta\sim\delta$ from \eqref{def:periodicity} and $f(z)\sim1/z^2$ since $0<z\leq z_h$. 

In our previous discussion, we took $\mu\sim1/\delta$, $M\sim1/\delta$ and $\beta\sim\delta$. 
Since the term with level $m$ in \eqref{res:large_gap_2pt} gives an exponential suppression, the leading order is given by $m=0$. 
Now we take all the $O(1)$ parameters to be $1$ and define 
\begin{align}
    P(\delta) = \frac{2}{\delta}\delta^{-2/\delta} \exp\left(-2\int_z^{\delta}\dint z \left(\sqrt{\frac{1}{z^2\delta^2}-\frac{1}{\delta^2}}-\frac{1}{z\delta}\right)\right) e^{-\pi/\delta}, 
\end{align}
which indicates the dependence of $\delta$ in \eqref{res:large_gap_2pt}. 
The integral of $z$ in the limit $z\to0$ gives 
\begin{align}
    \int_z^{\delta}\dint z \left(\sqrt{\frac{1}{z^2\delta^2}-\frac{1}{\delta^2}}-\frac{1}{z\delta}\right) &\sim \frac{1}{\delta} \left({\rm arctanh}\left(\sqrt{1-z^2}\right)-{\rm arctanh}\left(\sqrt{1-\delta^2}\right)+\log\left(\frac{z}{\delta}\right)\right) \nonumber \\
    &\sim \frac{1}{\delta}\left(\log\left(\frac{\delta}{z}\right)+\frac{1}{4}(\delta^2-z^2)\right). 
    \label{eqn:int_inside_expo}
\end{align}
Plugging \eqref{eqn:int_inside_expo} into $P(\delta)$, we will get 
\begin{align}
    P(\delta) &\sim \frac{2}{\delta} \delta^{-4/\delta} z^{2/\delta} e^{-(\delta^2-z^2)/2} e^{-\pi/\delta} \nonumber \\
    &\sim \frac{2}{\delta} \delta^{-4/\delta} z^{2/\delta} e^{-\pi/\delta}. 
    \label{eqn:asym_form_P}
\end{align}
If we fix the ratio $\delta/z\equiv\xi$ and send them to zero together, numerically we find that as long as $\xi\sim30$, $P(\delta)$ will be $\sim O(10^{-30})$ when $\delta\sim0.06$. 
The large gapped case is much suppressed as we expected, since inflation cannot effectively excite the largely gapped unparticles. 
\clearpage
\addcontentsline{toc}{section}{References}
\bibliographystyle{utphys}
{\linespread{1.075}
\bibliography{Refs}
}

\end{document}